\documentclass[8.5pt,twoside]{article}
\usepackage[super,sort&compress,comma]{natbib} 
\usepackage{times,mathptm}
\usepackage{sectsty}
\usepackage{balance} 
\usepackage[text={11.3cm,20.4cm},centering]{geometry} 

\usepackage{graphicx} 
\usepackage{lastpage}
\usepackage[format=plain,justification=raggedright,singlelinecheck=false,font=small,labelfont=bf,labelsep=space]{caption} 
\usepackage{fancyhdr}
\newcommand {\apjs} {\it Astrophys. J., Suppl. Ser.}
\newcommand {\apjl} {\it Astrophys. J. Lett.}
\newcommand {\aap} {\it Astron. Astrophys.}

\begin{document}

\thispagestyle{plain}
\fancypagestyle{plain}{
\fancyhead[L]{\includegraphics[height=8pt]{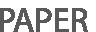}}
\fancyhead[C]{\hspace{-1cm}\includegraphics[height=15pt]{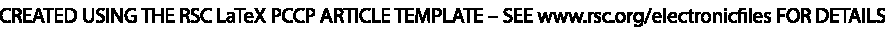}}
\fancyhead[R]{\includegraphics[height=10pt]{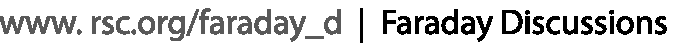}\vspace{-0.2cm}}
\renewcommand{\headrulewidth}{1pt}}
\renewcommand{\thefootnote}{\fnsymbol{footnote}}
\renewcommand\footnoterule{\vspace*{1pt}%
\hrule width 11.3cm height 0.4pt \vspace*{5pt}} 
\setcounter{secnumdepth}{5}

\makeatletter 
\renewcommand{\fnum@figure}{\textbf{Fig.~\thefigure~~}}
\def\subsubsection{\@startsection{subsubsection}{3}{10pt}{-1.25ex plus -1ex minus -.1ex}{0ex plus 0ex}{\normalsize\bf}} 
\def\paragraph{\@startsection{paragraph}{4}{10pt}{-1.25ex plus -1ex minus -.1ex}{0ex plus 0ex}{\normalsize\textit}} 
\renewcommand\@biblabel[1]{#1}            
\renewcommand\@makefntext[1]%
{\noindent\makebox[0pt][r]{\@thefnmark\,}#1}
\makeatother 
\sectionfont{\large}
\subsectionfont{\normalsize} 

\fancyfoot{}
\fancyfoot[LO,RE]{\vspace{-7pt}\includegraphics[height=8pt]{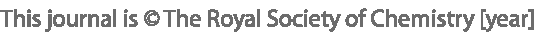}}
\fancyfoot[CO]{\vspace{-7pt}\hspace{5.9cm}\includegraphics[height=7pt]{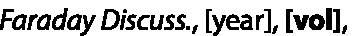}}
\fancyfoot[CE]{\vspace{-6.6pt}\hspace{-7.2cm}\includegraphics[height=7pt]{RF}}
\fancyfoot[RO]{\scriptsize{\sffamily{1--\pageref{LastPage} ~\textbar  \hspace{2pt}\thepage}}}
\fancyfoot[LE]{\scriptsize{\sffamily{\thepage~\textbar\hspace{3.3cm} 1--\pageref{LastPage}}}}
\fancyhead{}
\renewcommand{\headrulewidth}{1pt} 
\renewcommand{\footrulewidth}{1pt}
\setlength{\arrayrulewidth}{1pt}
\setlength{\columnsep}{6.5mm}
\setlength\bibsep{1pt}

\noindent\LARGE{\textbf{Astrochemistry of dust, ice and gas: introduction and overview}}
\vspace{0.6cm}

\noindent\large{\textbf{Ewine F. van Dishoeck\textit{$^{a,b}$}}}\vspace{0.5cm}

\noindent\textit{\small{\textbf{Received June 28, 2014; Accepted June 30, 2014\newline
Published as Faraday Discussions 2014, 168, 9}}}

\noindent \textbf{\small{DOI: 10.1039/c4fd00140k}}
\vspace{0.6cm}

\noindent \normalsize{A brief introduction and overview of the
  astrochemistry of dust, ice and gas and their interplay is
  presented. The importance of basic chemical physics studies of
  critical reactions is illustrated through a number of recent
  examples. Such studies have also triggered new insight into
  chemistry, illustrating how astronomy and chemistry can enhance each
  other. Much of the chemistry in star- and planet-forming regions is
  now thought to be driven by gas-grain chemistry rather than pure
  gas-phase chemistry, and a critical discussion of the state of such
  models is given. Recent developments in studies of diffuse clouds
  and PDRs, cold dense clouds, hot cores, protoplanetary disks and
  exoplanetary atmospheres are summarized, both for simple and more
  complex molecules, with links to papers presented in this volume. In
  spite of many lingering uncertainties, the future of astrochemistry
  is bright: new observational facilities promise major advances in
  our understanding of the journey of gas, ice and dust from clouds to
  planets.} \vspace{0.5cm}

\footnotetext{\textit{$^{a}$Leiden Observatory, Leiden University, P.O. Box 9513, 2300 RA Leiden, the Netherlands; E-mail: ewine@strw.leidenuniv.nl}}
\footnotetext{\textit{$^{b}$}Max-Planck-Institute f\"ur Extraterrestrische Physik, Giessenbachstrasse 1, Garching, 85748, Germany}

\section{Introduction}

The space between the stars is not empty, but is filled with a very
dilute gas, the interstellar medium (ISM). The ISM is far from
homogeneous and contains gas with temperatures ranging from more than
$10^6$ K down to 10 K and densities from as low as $10^{-4}$ particles
cm$^{-3}$ to more than $10^8$ cm$^{-3}$. The colder and denser
concentrations of the gas are called interstellar clouds, and this is
where molecules, dust and ice are detected.  Even at the upper range
of densities, interstellar clouds are still more tenuous than a
typical ultra-high vacuum laboratory experiment on Earth. Thus,
interstellar space provides a unique environment in which chemistry
can be studied under extreme conditions. From an astronomical
perspective, dense clouds are also important because they are the
nurseries of new generations of stars like our Sun and planets like
Jupiter or Earth.  This combination makes astrochemistry such a
fascinating research field, for both chemists and astronomers alike.

Traditional chemistry would predict that virtually no molecules are
formed at typical densities of 10$^4$ cm$^{-3}$ and temperatures of 10
K of dense molecular clouds, with 1000 times more hydrogen than any
other chemically interesting element. The detection of nearly 180
different species over the past 45 years (not counting isotopologs)
demonstrates the opposite: there is a very rich chemistry in space.
Molecules like CO and H$_2$O have even been detected in distant
galaxies out to high redshifts, when the Universe was less than 1 Gyr
old \citep{Weiss13}. Large Polycyclic Aromatic Hydrocarbon molecules
(PAHs) are also detected across the Universe
\citep{Tielens08}. Closer to home, an increasing variety of molecules
(no longer just the simplest ones) are found in luminous galaxies
undergoing bursts of star formation at a rate up to 100 times higher
than that in our own Milky Way \cite{Gao07,Martin11}.  In star-forming
clouds within our Galaxy, a myriad of species has been found, some of
which are prebiotic, i.e., molecules that are thought to be involved
in the processes leading to living organisms. Fullerenes such as
C$_{60}$ have also been identified \cite{Cami10}.
Finally, simple molecules including H$_2$O are being detected in the
atmospheres of giant exo-planets \cite{Seager10,Birkby13}.

Astrochemistry, also known as molecular astrophysics, is `the study of
the formation, destruction and excitation of molecules in astronomical
environments and their influence on the structure, dynamics and
evolution of astronomical objects' as stated by
\citet{Dalgarno08}. This definition includes not only the chemical
aspects of the field, but also the fact that molecules are excellent
diagnostics of the physical conditions and processes in the regions
where they reside. This sensitivity stems from the fact that both the
excitation and abundances of molecules are determined by collisions,
which in turn are sensitive to gas temperature and density as well as
to the radiation from nearby stars.

The main questions in the field of astrochemistry therefore include:
how, when and where are these molecules produced and excited?  How far
does this chemical complexity go? How are they cycled through the
various phases of stellar evolution, from birth to death? And, most
far-reaching, can they become part of new planetary systems and form
the building blocks for life elsewhere in the Universe?

The topic of this Symposium ---dust, ice and gas--- is at the heart of
these questions. This brief overview is aimed at providing background
information for non specialists on the conditions and ingredients in
interstellar space (\S 2), the tools that are used in astrochemistry
(telescopes, laboratory experiments, computer simulations) (\S 3), the
main characters of this symposium (dust, ice and gas) (\S 4), the
`play' (chemical processes) (\S 5), and the `plot' (evolution from
clouds to planets (\S 6--11), followed by some concluding remarks (\S
12).  There has been no shortage of excellent and much more detailed
(and partly overlapping) reviews of the field over the last few years,
see
\cite{Tielens13,vanDishoeck13,Herbst13,Caselli12rev,Bergin13,Herbst09}
for further information.

Throughout this talk the interdisciplinary aspect of this research
will be emphasized as a two-way street: astrochemistry needs basic
data on molecular spectroscopy and chemical processes, but also
inspires new chemical physics through studies of different classes of
molecules and reactions that are normally not considered on
Earth. Indeed, astrochemistry is a `blending of astronomy and
chemistry in which each area enriches the other in a mutually
stimulating interaction' \cite{Dalgarno08}.

\begin{figure}[h]
\centering
 \includegraphics[width=9cm]{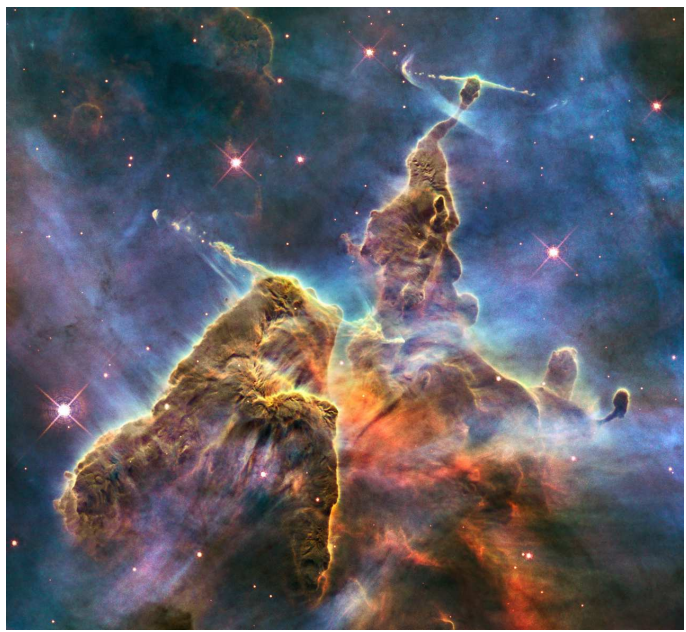}
\caption{Blow-up of the Hubble Space Telescope optical image of the
  Carina nebula, showing dark molecular clouds as well as more diffuse
  ionized gas between the stars. The clouds are dark at visible
  wavelengths due to extinction by dust grains; young stars are
  embedded within them. In the colder regions, the dust particles are
  covered by ice mantles. Gas-phase molecules are present throughout the
  dark clouds.  The distance to the cloud is 2.3 kpc, and the image
  covers a region of 0.94 pc in size. Colors: forbidden transitions of
  [O III] (blue), [N II] (green), and [S II] (red), together with
  hydrogen H$\alpha$ (green). Credit: NASA/ESA/M. Livio. }
  \label{fig:carina}
\end{figure}

\section{The setting: interstellar clouds}

\subsection{Images of clouds}

Figure~\ref{fig:carina} presents an {\it Hubble Space Telescope} image
of a small part of the Carina star-forming cloud. The
colorful nebulae are due to ionized gas at $10^4$ K that emits
brightly at optical wavelengths and consists of atomic hydrogen
recombination lines (primarily H$\alpha$) and
forbidden electronic transitions of atoms such as [O III], [N II] and [S II].
The dark areas are the dense molecular clouds which
contain $\sim$0.1 $\mu$m-sized particles of dust composed of silicates
and carbonaceous compounds which absorb and scatter the light. They
also shield molecules from the dissociating ultraviolet radiation
emitted by nearby stars.  The dense cloud complexes can be quite large
(tens of parsec\footnote{1 parsec (pc)=3.1$\times 10^{18}$ cm = 3.26
light year} across) and massive (up to 10$^5$ solar masses\footnote{1
M$_\odot$=$2.0\times 10^{33}$ gr}) but the process of star formation is
quite inefficient and only a few~$\%$ of the mass of the cloud is
turned into stars.

The UV radiation absorbed by dust grains is converted to heat, raising
their temperatures from $\sim$10 K in the most shielded regions to
$\sim$50 K in more exposed gas. This heat is then radiated away
through thermal emission in the far-infrared part of the
spectrum as observed for example with the {\it Herschel Space Observatory}.

\begin{figure}[h]
\centering
\includegraphics[width=10cm]{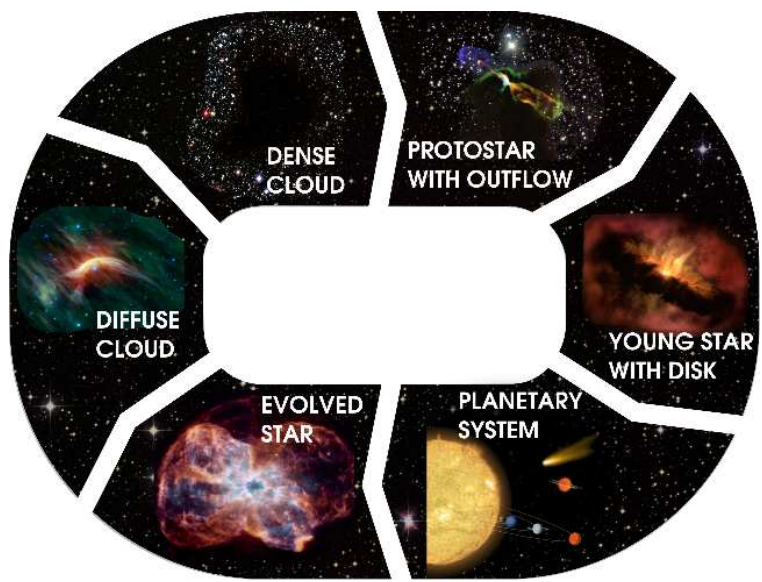}
\caption{Various stages in the lifecycle of gas, dust and ice in interstellar and circumstellar clouds. Figure by M. Persson; 
credit NASA/ESA/ESO/ALMA for pictures. }
  \label{fig:cycle}
\end{figure}

\subsection{Birth and death of stars}

Figure~\ref{fig:cycle} illustrates the cycle of material from clouds
to stars and planets, and ultimately back to the interstellar medium.
Diffuse clouds are low-density clouds ($\sim 100$ cm$^{-3}$) in which
UV radiation can penetrate and destroy molecules. They are usually
transient structures in the interstellar medium. Dense molecular
clouds can be stable for millions of years due to support by
turbulence and magnetic fields, but eventually gravity takes over and
the densest part of the cloud collapses to form a new star. In the
standard scenario, the collapse occurs inside out so the protostar at
the center of the cloud continues to grow as material from the
envelope accretes onto the star. Because of angular momentum
conservation, the material that falls in at later times ends up in a
disk around the star where the gas is in Keplerian
rotation. Further accretion of material takes place in the innermost
part of the disk in magneto-hydrodynamically (MHD) mediated funnel
flows onto the star.  The MHD processes near the disk-star boundary
also result in jets and winds which can escape in a direction
perpendicular to the disk. When they interact with the surrounding
envelope and cloud, they create shocks and entrain material in bipolar
outflows.

With time ($\sim$1 Myr=$10^6$yr), the opening angle of the wind
increases and the envelope is gradually dispersed, revealing a young
pre-main sequence star surrounded by a so-called protoplanetary disk.
These disks are about the same size as the Solar System ($\sim$100
AU\footnote{1 AU = distance Sun-Earth = 1.5$\times 10^{13}$ cm}) and
contain a few Jupiter masses of gas and dust (1 M$_{\rm
  Jup}$=$1.9\times 10^{30}$ gr or about 0.1\% of the mass of the Sun).
Here densities can be as high as $10^{13}$ cm$^{-3}$ in the inner
regions of disks. The increased collision rates lead to gradual
coagulation of dust grains to form pebbles, rocks and planetesimals,
although the precise mechanisms are not yet understood \citep{Blum08}.
The large particles settle to the midplane, where they can form
kilometer-sized objects that interact gravitationally to form
(proto)planets and eventually a full planetary system (up to 100 Myr).
Comets and asteroids are remnant planetesimals that did not end up in
one of the planets and were scattered and preserved in the cold outer
regions of our Solar system. If they have not been heated during their
lifetime, their composition reflects the conditions during the solar
system formation.

The young star eventually becomes hot enough for nuclear fusion of hydrogen to
ignite, at which stage it enters the stable main-sequence phase for
several Gyr. Photospheric temperatures of low-mass stars are typically
$T_*$=3000--5000 K so they emit most of their radiation at visible (as
our Sun) and near-infrared wavelengths. High-mass stars have $T_*$ up
to 40,000 K so their Planck spectra peak at far-ultraviolet
wavelengths, which are much more effective at destroying molecules.

At the end of the stellar life cycle, the nuclear fuel is exhausted,
causing the low-mass stars to become larger and loose part of their
mass (Fig.~\ref{fig:cycle}). These evolved Asymptotic Giant Branch
(AGB) stars are surrounded by circumstellar envelopes, i.e., dense
shells of molecular material driven by radiation
pressure. Temperatures are high, 2000--3000 K near the stellar
photosphere, dropping with radius down to 10 K at the outer edge.
Because of the dense and warm conditions, they are a rich source of
molecules. Such conditions are also favorable for the condensation of
dust grains.  Once the envelopes have been dissipated, the
central star becomes very hot and emits copious UV radiation,
illuminating the surrounding remnant gas producing a so-called `planetary
nebula' (not to be confused with a (proto)planetary disk). Massive
stars explode after a much shorter lifetime ($\sim$10 Myr) as
supernovae with their cores ending up as black holes or neutron stars.

\subsection{UV radiation and cosmic rays}

The UV radiation impinging on an interstellar cloud is often
approximated by a scaling factor called $G_o$ or $I_{\rm UV}$ with
respect to the average radiation produced by all stars in the solar
neighborhood from all directions. This interstellar radiation field
(ISRF) $I_0$ has been estimated by \citet{Habing68} and \citet{Draine78} to
have an intensity of about $10^8$ photons cm$^{-2}$ s$^{-1}$, with
a relatively flat spectrum between the threshold at 912 \AA\ and 2000
\AA.  Any photon with energies greater than 13.6 eV (wavelength $<$ 912
\AA) is absorbed by atomic hydrogen outside the cloud and therefore does not
affect the chemistry.

Dust particles absorb UV radiation and thereby shield molecules deeper
into the cloud from the harshest radiation.  The extinction at visual
wavelengths, $A_V$, is defined as $ 1.086 \times \tau_d$ at 5500 \AA,
wit $\tau_d$ the optical depth of the dust.  The intensity
decreases as $I_{5500}=I_0 10^{-0.4 A_V}$ with depth into a cloud.  At
UV wavelengths, the decline is much steeper but depends on the grain
properties such as size, composition, shape and scattering
characteristics \citep{Roberge91}.

Cosmic ray particles are another important ingredient of interstellar
clouds. These are highly energetic atomic nuclei with $>$MeV
energies. Cosmic rays penetrate much deeper into clouds than UV
radiation and ionize a small fraction of atomic and molecular hydrogen
needed to kick-start the gas-phase chemistry.

Cosmic rays also maintain a low level of UV radiation deep inside
dense clouds \citep{Prasad83}: the ionization of H and H$_2$ produces
energetic secondary electrons which excite H$_2$ into the
B$^1\Sigma_u^+$ and C$^1\Pi_u$ electronic states, which subsequently
decay through spontaneous emission in the Lyman and Werner bands. The
resulting UV spectrum consists of discrete lines and a weak continuum
in the 900--1700 \AA\ range \citep{Gredel89}. The flux of internally
generated UV photons is typically $10^4$ photons cm$^{-2}$ s$^{-1}$
but depends on the energy distribution of the cosmic rays (see Fig.~4
of \citealt{Shen04}) and the grain properties.

Interstellar clouds are largely neutral, since hydrogen-ionizing
photons have been absorbed in the more diffuse surrounding gas. Of the
major elements, only carbon can be ionized by the ISRF because its
first ionization potential is less than 13.6 eV. 
Since the abundance of gas-phase carbon with respect to hydrogen is
about $10^{-4}$ (see Table~\ref{tab:abundances}), this sets the
maximum electron fraction in the cloud to about $10^{-4}$.
With depth into the cloud, the ionized carbon is converted into
neutral atomic and molecular form. Around $A_V$=5 mag, cosmic rays
take over as the main ionizing agent at a rate denoted by $\zeta$ in
s$^{-1}$.  The resulting ionization fraction depends on the detailed
chemistry and grain physics but is typically $ 10^{-7}$ or lower, and
scales as $(n/\zeta)^{-1/2}$.

\section{The tools: telescopes, laboratory}

\subsection{Importance of multi-wavelength observations}

The energy levels of a molecule are quantized into electronic,
vibration and rotation states, with decreasing energy difference
between two neighboring levels of the same type. Electronic
transitions typically occur at optical and UV wavelengths, whereas
those between two vibrational levels within the same electronic state
take place at infrared wavelengths.  Rotational transitions within a
given electronic and vibrational state are found at (sub)millimeter
and far-infrared wavelengths. 

The advantages of optical and UV spectroscopy are that a number of key
species, most notably H$_2$ and atoms, can be observed directly. The
oscillator strengths of the transitions involved are large, so even
minor species can be detected.  The drawback is that only diffuse
clouds and translucent clouds with less than a few mag of extinction
can be observed along the line of sight toward a background source,
since short wavelength photons do not penetrate thicker clouds.

The main advantage of infrared spectroscopy is that not just gases but
also solids can be observed, both in absorption and emission. For the
simplest case of an harmonic oscillator, the energy levels are given by
$\omega_e ({\rm v} + 1/2)$ where $\omega_e$ is the vibrational
frequency and v the vibrational quantum number. Note the zero-point
vibrational energy for v=0, which has important chemical consequences
in cold dark clouds. The strongest vibrational bands of ices,
silicates, oxides and PAHs occur at mid- and far-infrared wavelengths
\citep{vanDishoeck04}. Some important gas-phase molecules without a
permanent dipole moment, such as H$_3^+$, CH$_4$, C$_2$H$_2$ and
CO$_2$, are also only observed through their vibrational
bands. Disadvantages are relatively low spectral resolving power (up
to $R=\lambda/\Delta \lambda =10^5$) and moderate oscillator
strengths, which means that in practice only a handful of molecules
are detected at infrared wavelengths.

The bulk of the interstellar molecules have deen discovered at
millimeter (mm) wavelengths. For the simplest case of a linear
molecule, the rotational energy levels scale as $B J(J+1)$ where $J$
is the rotational quantum number and $B$ the rotational constant,
which is inversely proportional to the reduced mass of the
molecule. Thus, the transitions of light molecules such as hydrides
occur at higher frequencies (THz) than those of heavy molecules. The
lowest levels have energies $E_u/k_B$ (with $k_B$=Boltzmann constant)
that range from a few to tens of K, and are thus readily excited at
typical dense cloud conditions. The advantage of mm observations is
high sensitivity to low abundance molecules (down to $10^{-11}$ with
respect to hydrogen) and the fact that the emission can be mapped so
that one is not limited to a single line of sight.

\subsection{Telescopes}

Progress in astronomy is very much driven by large telescopes equipped
with highly sensitive detectors that allow investigation of clouds at
a variety of wavelengths across the electromagnetic spectrum, each of
which tells a different part of the story. In the last decade
astrochemistry has been very fortunate to have had access to a number of
new powerful telescopes with both imaging and high resolution
spectroscopic instruments, the latter crucial for observing
molecules. In particular, the {\it Herschel Space Observatory} was the
largest astronomical telescope in space (3.5m diameter) operative from
mid-2009 to mid-2013. Specifically, the HIFI instrument provided very high
resolution ($R$ up to $10^7$) heterodyne spectroscopy covering the
490--1250 GHz (600--240 $\mu$m) and 1410--1910 GHz (210--157 $\mu$m)
bands for a single pixel on the sky \citep{deGraauw10}.  
At mid-infrared wavelengths, the {\it Spitzer
Space Telescope} had a low-resolution ($R$=50--600) spectrometer at
5--40 $\mu$m from 2003--2009, which was particularly powerful to study
dust features and ices in large numbers of sources due to its high
sensitivity, following on the pioneering results from the {\it
  Infrared Space Observatory} (ISO) \citep{vanDishoeck04}.

On the ground there have also been significant advances.  The European
Southern Observatory (ESO) -Very Large Telescope (VLT) and the Keck
and Subaru Observatories provide the most powerful collection of
8--10m optical-infrared telescopes on Earth. At millimeter wavelengths
the single-dish 10m Caltech Submillimeter Observatory, 15m {James
  Clerk Maxwell Telescope}, the 45m Nobeyama dish, the IRAM 30m
telescope in Spain and the 12m Atacama Pathfinder Experiment (APEX)
have opened up the (sub)millimeter regime and have been workhorses for
astrochemistry over the past 30 years. To obtain higher spatial
resolution, pioneering millimeter interferometers such as the IRAM
Plateau de Bure and the CARMA array have provided a first
glimpse. Astronomers usually express spatial resolution and scales in
arcsec ($''$) \footnote{One arcsec ($''$) is 1/3600 of a degree or
  1/206265 of a radian ($\pi$/180/3600).  At a distance of
  $n\times$100 pc, 1 arcsec corresponds to $n\times$100 AU diameter,
  or $n\times$ 100$\times$ the Sun-Earth distance.}. The {\it Atacama
  Large Millimeter/submillimeter Array} (ALMA), located in north Chile
on the Chajnantor plateau at an altitude of 5000 meter, became
operational in 2011 and will be transformational for the field.  It
consists of 54 $\times$12m and 12$\times$7m antennas which together
can form images with a sharpness ranging from $\sim$0.01$''$ to a few
$''$ over the frequency range from 84 to 900 GHz, and with a
sensitivity up to two orders of magnitude higher than any previous
instrument.

\subsection{Laboratory experiments}
 
The most basic information that astronomers need from laboratory
experiments is spectroscopy from UV to millimeter wavelengths, to
identify the sharp lines and broad bands observed toward astronomical
sources.  Techniques range from classical absorption set-ups to the
use of cavity ringdown spectroscopy to increase the sensitivity by
orders of magnitude \citep{Linnartz10}.  Transition frequencies and
strengths of (sub)mm transitions are summarized in the Jet Propulsion
Laboratory catalog (JPL) \citep{Pickett98} and the Cologne Database
for Molecular Spectroscopy (CDMS) \citep{Muller01,Muller05}
\footnote{{\tt spec.jpl.nasa.gov} and {\tt
    www.astro.uni-koeln.de/cdms/catalog}}.  Databases for vibrational
transitions at infrared wavelengths include the HITRAN database
\citep{Hitran09} and the EXOMOL line lists\footnote{{\tt
    www.cfa.harvard.edu/hitran} and {\tt www.exomol.com}}.

The spectroscopy of large samples of PAHs has been determined with
matrix-isolation studies, with selected PAHs also measured in the gas
phase \cite{Boersma14}. Spectroscopic data bases of solids continue to
grow and include the Heidelberg-Jena-St.\ Petersburg database of
optical constants for silicates \citep{Henning10}, various databases
for ices \citep{Hudgins93,Linnartz11,Hudson14,Munoz14} and
carbonaceous material \citep{Mennella10,Munoz06}. THz spectroscopy of
ices is just starting, as reported by Ioppolo et al.\ in this volume.

The next step is to determine the rates for the various reactions that
are expected to form and destroy molecules under space conditions are
needed.  Laboratory experiments for gas-phase processes have recently
been summarized by \citet{Smith11}.  Developments include measurements
and theory of gaseous neutral-neutral rate coefficients at low
temperatures using techniques such as CRESU \citep{Chastaing01} or
crossed molecular beams \citep{Morales11}, branching ratios for
dissociative recombination \citep{Geppert13}, and rates for
photodissociation of molecules exposed to different radiation fields
\citep{Mainz}.

For ices, the combination of observations and associated modeling have
stimulated the new field of solid-state laboratory astrochemistry, in
which modern surface science techniques at ultra-high vacuum
conditions are used to quantitatively study various chemical
processes. A recent summary of activities across the world is given by
\citet{Allodi13}.  Temperature programmed desorption experiments now
routinely probe thermal desorption of both pure and mixed ices,
providing binding energies that can be used in models
\citep{Collings04,Bisschop06,Munoz14desorption}. Surface chemistry
experiments have demonstrated that solid CH$_3$OH
\citep{Watanabe02,Hidaka04,Fuchs09} and H$_2$O
\citep{Ioppolo08,Miyauchi08,Dulieu10,Cuppen10,Accolla13} can indeed
form at temperatures as low as 10 K, using beams of atomic H to
bombard solid CO, O, O$_2$ and O$_3$, thereby confirming routes that
were postulated more than 30 years ago by \citet{Tielens82}.  The process
of photodesorption has been measured quantitatively and found to be
much more efficient than previously thought
\citep{Oberg07,Oberg09h2o}, and to depend strongly on the wavelength
of the incident radiation field \citep{Fayolle11,Bertin13}.  Details
of the formation of complex organic molecules by UV irradiation and by
high-energy particle bombardment
\citep{Moore01,Elsila07,Oberg09meth,Kaiser11,Gerakines12,Islam14}
continue to be elucidated by a variety of experiments.
Finally, the techniques to analyze meteoritic and cometary material in
the laboratory have improved enormously in the last decade, and now
allow studies of samples on submicrometer scale using techniques such
as ultra-L$^2$MS, nano-SIMS and NMR \cite{Cody11}.

\subsection{Theory and Computation}

Theoretical studies can be equally important as laboratory experiments
in providing information on astrochemically relevant questions.  The
most important example is that of collisional rate coefficients,
needed to determine the excitation of interstellar molecules. Although
laboratory experiments can provide some selected data, only theory can
provide the thousands of state-to-state cross sections as function of
temperature needed in astronomy.  The process consists of two
steps. First, {\it ab initio} quantum chemical methods are used to
compute the multi-dimensional potential energy surface of the molecule
and its collider (usually H$_2$). This surface is then fitted to a
convenient functional or numerical form for use in the second step,
the dynamics of the nuclei.  Inelastic scattering calculations need to
be computed over a large range of collision energies. Full quantum
calculations are used at the lowest energies whereas quasi-classical
methods are often employed at the highest energies. A large number of
systems has been studied over the past two decades, as summarized in a
number of reviews \citep{Schoier05,vanderTak11,Dubernet13}.

Theoretical studies are also important for computing cross sections
and rates for other processes. A prime example is photodissociation
rates of small radicals and ions through calculation of the excited
electronic states and their transion dipole moments
\citep{vanDishoeck06photo,vanHemert08,Stancil13}. Other examples
include radiative association reactions \citep{Lin13}, rates and
barriers for neutral-neutral reactions \citep{Woon97,Herbst00},
state-selective reactions \citep{Li13oh}, and the structure
\citep{Woon09} and infrared spectroscopy \citep{Ricca13} of large
molecules.

Molecular dynamics techniques can also be extended to solid-state
processes, as demonstrated by the detailed study of the
photodissociation and photodesorption of water ice and its isotopologs
\citep{Andersson06,Arasa10,Koning13}. Another recent example is the
attempt to model the formation of solid CO$_2$ ice \citep{Arasa13}.

\section{The main players: dust, ice and gas}

\subsection{The ingredients: elemental abundances}
\label{sect:abundances}

The astronomers' view of the periodic table is fairly restricted. The
local Universe consists primarily of 90\% hydrogen with about 8\%
helium by number. All other elements are called `metals' by
astronomers even if they are obviously not a metal in a chemical
sense.  The next most abundant elements are oxygen, carbon, and
nitrogen at abundances of only about 4.9, 2.7 and 0.7 $\times 10^{-4}$
that of hydrogen.  The abundances summarized in
Table~\ref{tab:abundances} are those derived for the solar photosphere
by \citet{Asplund09}. These abundances are thought to apply to the
interstellar medium as well, although small variations are possible
\citep{Przybilla08}.

\begin{table}[h]
\small
\centering
  \caption{Solar elemental abundances}
  \label{tab:abundances}
  \begin{tabular}{l c l c} 
    \hline
    Element & Abundance & Element & Abundance \\
    \hline
    H & 1.00               & Mg & $4.0\times 10^{-5}$ \\
    He & 0.085             & Al & $2.8\times 10^{-6}$ \\
    C & $2.7\times 10^{-4}$ & Si & $3.2\times 10^{-5}$ \\
    N & $6.8\times 10^{-5}$ & S  & $1.3\times 10^{-5}$ \\
    O & $4.9\times 10^{-4}$ & P  & $2.6\times 10^{-7}$ \\
    Na & $1.7\times 10^{-6}$ & Fe & $3.2\times 10^{-5}$ \\
    \hline
  \end{tabular}
\end{table}

\subsection{Interstellar dust}
\label{sect:dust}

The size, composition, shape and other properties of interstellar dust
have been studied through a wide variety of observational techniques,
such as extinction, reflection and emission measurements, including
polarization techniques, as well as spectroscopy (see reviews and
books by \citealt{Draine03,Draine11} and \citealt{Tielens05}). It is
clear that the bulk of the dust grains consist of a mix of amorphous
silicates and carbonaceous material, which locks up nearly 100\% of
the Si, Mg and Fe, $\sim 30$\% of the oxygen and about 70\% of the
available carbon. These materials are called `refractory' since they
do not vaporize until temperatures well above 1200 K. They are
therefore not available for the `volatile' gas or ice chemistry. Other
types of solids such as iron oxides, carbides, sulfides or metallic
iron are also present, but are at most a minor component.

Interstellar grains have a distribution of sizes $a$ following roughly
$a^{-3.5}$. This means that most of the surface area for chemistry is
in grains that are smaller than the typcal 0.1 $\mu$m-sized grains
that dominate the mass.  These smaller grains, down to 0.001 $\mu$m or
less, dictate the absorption and scattering of UV radiation. In dense
cores and circumstellar disks, grain growth up to sizes of a few
$\mu$m to a few cm has been found \citep{Oliveira11,Testi14}.  At low
temperatures, grains in dense cores are covered with a thick ice
mantle consisting primarily of H$_2$O ice. If the bulk of the
`volatile' oxygen is locked up in ice, a typical 0.1 $\mu$m-sized
silicate grain core is surrounded by about 100 monolayers of water
ice.

Polarization observations show that interstelar grains are not
spherical but elongated. Whether or not they have a highly
irregular and/or porous structure, such as seen in the larger
Interplanetary Dust Particles (see Wikipedia for images), is still
under debate. The bulk of the surface sites are suited for
physisorption of atoms and molecules from the gas, but there are
likely also some chemisorption sites available. Indeed, \citet{Jones13} 
(see also this conference) has suggested that grains in diffuse
clouds are coated by a layer of amorphous hydrocarbon material a-C(:H)
which will change the surface site properties.  

More generally, interstellar grains are not thought to have an active
role as catalysts in the chemical sense, in which surface molecules
participate in promoting the reaction. Rather, they provide a
reservoir where atoms and molecules from the gas can be stored and
brought together for a much longer period than possible in the
gas. They thus enable reactions with activation barriers that are too
slow in the gas such as the hydrogenation of atomic O, C and N. Also,
they act as a third body that absorbs the binding energy of a newly
formed molecule, thereby stabilizing it before it can dissociate
again.

\begin{figure}[h]
\centering
  \includegraphics[width=10cm]{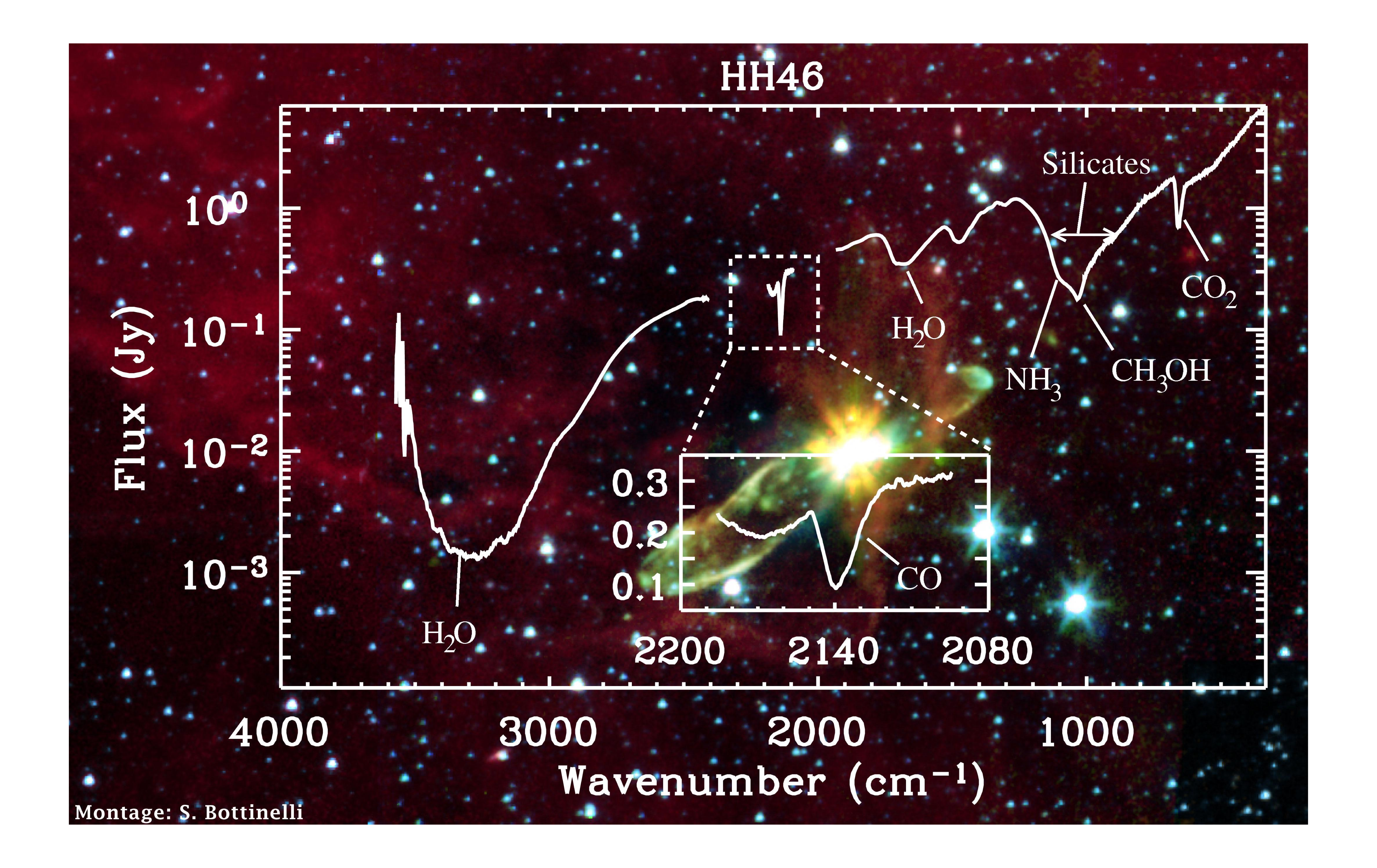}
\caption{Detection of ices toward the low-mass protostar HH46 IRS
  using {\it Spitzer} data at 5--20 $\mu$m and VLT-ISAAC data at 2--5
  $\mu$m \citep{Boogert04,Boogert08}. The strong solid CO$_2$
  stretching band at 4.3 $\mu$m is missing since it cannot be observed
  from the ground.  The insert shows a blow-up of the strong solid CO band;
  the weaker feature at 2167 cm$^{-1}$ is due to OCN$^-$. Montage by
  S.\ Bottinelli. Background: {\it Spitzer} composite 3 (blue, stars),
  4.5 (green, shocked H$_2$), and 8 $\mu$m (red, PAHs) image, showing
  the embedded protostar with its outflow. Credit
  NASA/ESA/A. Noriego-Crespo.}
  \label{fig:icespectrum}
\end{figure}

\subsection{Interstellar ices}
\label{sect:ices}

Ices are primarily observed in absorption against bright mid-infrared
sources, either embedded in the cloud or behind it
(Fig.~\ref{fig:icespectrum}). The Kuiper Airborne Observatory and
especially ISO opened up the field of infrared spectroscopy unhindered
by the Earth's atmosphere, allowing the first full inventory of
interstellar ices.  The dominant ice species are simple molecules,
H$_2$O, CO, CO$_2$, CH$_4$, NH$_3$ and CH$_3$OH, whose presence has
been firmly identified thanks to comparison with laboratory
spectroscopy \citep{Hudgins93,Linnartz11}.  These molecules are
precisely the species predicted to be produced by hydrogenation and
oxidation of the dominant atoms (O, C and N) and molecules (CO)
arriving from the gas on the grains at low temperatures
\citep{Tielens82}.  One ion, OCN$^-$, has also been convincingly
detected along many lines of sight
\citep{Novozamsky01,vanBroekhuizen05}. More recent {\it Spitzer}
surveys have extended such studies to large samples of low-mass YSOs
(Fig.~\ref{fig:icespectrum}). An important conclusion is that these
sources have a remarkably similar overall composition as their
high-mass counterparts within factors of two \citep{Oberg11}.  This
implies that the formation mechanism of these `zeroth generation' ice
species is robust and universal across the Galaxy.

\begin{figure}[h]
\centering
\includegraphics[width=10cm]{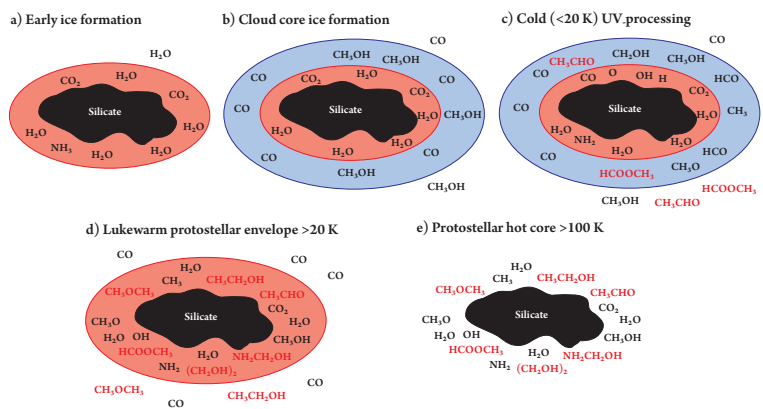}
\caption{Proposed evolution of ices during star formation and
  formation of complex molecules. Pink indicates an H$_2$O-dominated
  (`polar') ice and blue a CO-dominated (`non-polar') ice. Early
  during cloud formation (a) an H$_2$O-rich ice forms, with minor
  amounts of CH$_4$ and NH$_3$ through hydrogenation of O, C and
  N. Once a critical density is reached CO freezes out
  catastrophically (b), providing reactants for CH$_3$OH ice formation
  through hydrogenation of CO (0$^{\rm th}$ generation complex organic
  molecules). In the cold outer envelope (c), photoprocessing of the
  CO-rich ice results in the production of, e.g., HCOOCH$_3$ (1$^{\rm
    st}$ generation).  Closer to the protostar (d), following
  sublimation of CO, other complex molecules become abundant. Finally,
  all ices desorb thermally close to the protostar $>$100 K (e)
  (2$^{\rm nd}$ generation).  Figure from \citet{Oberg10b1},
reproduced with permission.}
  \label{fig:iceevolution}
\end{figure}

Interstellar ices are known to have a layered structure surrounding
the silicate or carbonaceous cores: a water-rich and a water-poor
layer (also called `polar' and `apolar' or `non-polar' ices) (see
Fig.~\ref{fig:iceevolution}). The observational evidence for different
layers comes from the shapes of various ice bands, in particular the
high-quality line profiles of solid CO
\citep{Tielens91,Pontoppidan03}: a CO molecule surrounded by a H$_2$O
molecule has a slightly different vibrational constant than a CO
molecule embedded in CO itself, and these differences can be readily
distinguished.  The water-rich layer is thought to form early in the
evolution of a cloud by hydrogenation of atomic O, once the extinction
is a few mag \citep{Whittet13}. The bulk of solid CH$_4$ and NH$_3$ is
likely also formed at this stage. In contrast, the water-poor, CO-rich
layer is observed to form in much denser gas (typically $>10^5$
cm$^{-3}$) when the freeze-out time, $\tau_{fo}\approx2\times
10^9/n_{\rm H}$ yr, has become so short compared with the lifetime of
the core that the bulk of the heavy elements are removed from the gas
`catastrophically' \citep{Caselli99,Pontoppidan06,Bergin02}.

\subsection{Interstellar gas}

About 180 different molecules have been detected in interstellar
space, not counting isotopologs\footnote{See {\tt
    www.astro.uni-koeln.de/cdms/molecules} or {\tt
    www.astrochemystry.org} for summaries}. The bulk of the molecules
are organic, i.e., they contain at least one carbon atom together with
(usually) at least one hydrogen atom. Several molecules were first
found in space before they were ever synthesized in a laboratory on
Earth. This includes ions such as HCO$^+$ and N$_2$H$^+$ and radicals
such as the long carbon chains HC$_9$N and C$_6$H. Only a few ring
molecules have so far been found, with c-C$_3$H$_2$ the best known
case. This does not necessarily imply that ring molecules are less
abundant than linear chains: it may simply be an observational effect
since the energy level structure and hence the partition function are
less favorable for cyclic molecules. Several of the organic molcules
are isomers, i.e., molecules with the same atomic constituents but
different structures. Well known examples are HCN and HNC, methyl
formate (HCOOCH$_3$), acetic acid (CH$_3$COOH)\citep{Remijan03} and
glycolaldehyde (HCOCH$_2$OH), and most recently HNCO versus HCNO, HOCN
and HONC \citep{Quan10}.

The saturated organic molecules are found primarily in hot cores
associated with star formation. In contrast, the unsaturated long
carbon-chain molecules are found predominantly in cold dark clouds
prior to star formation.  Molecules involving metals such as MgCN, AlOH,
and FeCN and phospor-containing molecules like HCP are primarily identified in
the envelopes around evolved stars.

The rate of detections is still on average 3 new molecules per
year. Recent highlights include the detection of new hydrides, such as
H$_2$O$^+$ \citep{Ossenkopf10,Benz10}, HCl$^+$ \citep{DeLuca12},
H$_2$Cl$^+$ \citep{Lis10chloor}, SH$^+$ \citep{Menten11} and SH
\citep{Neufeld12sh}.  Particularly exciting is the recent discovery of
the first interstellar noble gas molecule, $^{36}$ArH$^+$, in the gas
associated with the Crab nebula \citep{Barlow13}, the remnant of a
supernova that exploded in 1054 AD. This identification was
facilitated by the realization that in space, $^{36}$Ar is more
abundant than the $^{40}$Ar isotope commonly found on
Earth. $^{36}$ArH$^+$ is now also seen in diffuse clouds along the
line of sight to the Galactic Center and other distant sources
\citep{Schilke14}.

At the opposite end of the size range of interstellar molecules,
C$_{60}$ and C$_{70}$ have recently been identified in a young
planetary nebula through infrared spectroscopy \citep{Cami10}. Other
large molecules such as PAHs are also inferred to be present based on
the ubiquitous set of mid-infrared bands that are observed throughout
the Universe and which can be ascribed to aromatic C=C and C-H bonds
\citep{Tielens08}. However, in contrast with all other molecules
mentioned above, no individual PAH molecule has yet been identified;
likely a complex mix of PAH species makes up the observed features
\citep{Rosenberg14}.

\section{Interstellar chemistry}
\label{sect:chemistry}

\begin{table}[tb]
\caption{Types of Molecular Processes}
\label{tab:reactions}
{\begin{tabular}{llll}
\hline
\\ [1pt]
& {\it Bond Formation Processes} && Typical rate \\ 
&                                && coefficient (cm$^3$ s$^{-1}$) \\ [10pt]
&Radiative association & X + Y $\to$ XY + h$\nu$  & $10^{-17} - 10^{-14}$ \\
&Grain surface formation & X + Y:$g$ $\to$ XY + $g$  & $\sim 10^{-17}$ \\
&Associative detachment & X$^-$ + Y $\to$ XY + e & $\sim 10^{-9}$ \\ [10pt]
& {\it Bond Destruction Processes} & \\ [10pt]
&Photodissociation & XY + h$\nu$ $\to$ X + Y  & $10^{-10} - 10^{-8}$ s$^{-1}$ \\
&Dissociative recombination & XY$^+$ + e $\to$ X + Y  & $10^{-7} - 10^{-6}$\\
&Collisional dissociation & XY + M $\to$ X + Y + M & $\sim 10^{-26}$ cm$^6$ 
  s$^{-1}$ \\ [10pt]
& {\it Bond Rearrangement Processes} \\ [10pt]
&Ion--molecule exchange  & X$^+$ + YZ $\to$ XY$^+$ + Z & $10^{-9} - 10^{-8}$ \\
&Charge--transfer        & X$^+$ + YZ $\to$ X + YZ$^+$ & $10^{-9}$\\
&Neutral--neutral        & X + YZ $\to$ XY + Z   & $10^{-11} - 10^{-9}$  \\ [3pt]
\hline
\end{tabular}}
\end{table}

\subsection{Basic types of processes}
\label{sect:processes}

Table~\ref{tab:reactions} summarizes the basic types of reactions in
space. Because of the low densities, only two-body processes are
thought to be important.  The rate of a reaction between species X
and Y is given by $k n$(X)$n$(Y) in cm$^{-3}$ s$^{-1}$, where
$k$ is the reaction rate coefficient in cm$^3$ s$^{-1}$ and $n$
is the concentration in cm$^{-3}$.  Three-body reactions only become
significant at densities above $10^{13}$ cm$^{-3}$ such as encountered
in the atmospheres of AGB stars, the inner midplanes of protoplanetary
disks, and in the formation of the first stars in the Universe.

\subsubsection{Formation.} There are two basic processes by which molecular
bonds can be formed.  The first one is radiative association of atoms
or molecules, in which the binding energy of the new molecule is
carried away through the emission of photons.  The second process
involves formation on the surfaces of grains, in which the dust
particle accomodates the released energy (see
\S~\ref{sect:gasgrain}). Both of these processes are intrinsically
slow. For radiative association, typical timescales for infrared emission
are $10^{-3}$ s$^{-1}$, about 10 orders of magnitude slower than 
collision timescales of $10^{-13}$ s$^{-1}$ making the process
intrinsically very inefficient.  Only for specific cases, such as C$^+$ +
H$_2$ $\to$ CH$_2^+$ + h$\nu$, or for large molecules are the rate
coefficients increased by a few orders of magnitude. Radiative
association is so slow that it is very difficult to measure properly
in a laboratory on Earth, where three-body processes often dominate
\citep{Gerlich93}.

A third process, called associative detachment, is fast but requires
the presence of negative ions, the formation of which is the rate
limiting step. It plays a minor role in dense cloud chemistry but is
important, for example, in the chemistry of the early Universe.

\subsubsection{Destruction.} Diffuse clouds are permeated by
intense ultraviolet radiation, which destroys molecular bonds
through the process of photodissociation
\citep{vanDishoeck88photo,vanDishoeck11pd}. If the molecule has
dissociative excited electronic states below 13.6 eV,
photodissociation is very rapid, with typical molecular lifetimes of
only $\sim$100--1000 yr in the standard ISRF. Photodissociation can
also take place by absorption into excited states that are initially
bound, but which are subsequently either pre-dissociated or decay
through emission of photons into the vibrational continuum of the
ground state. The photodissociation of three of the most important
interstellar molecules, H$_2$ \citep{Abgrall00}, CO \citep{Visser09}
and N$_2$ \citep{Li13,Heays14}, is controlled by these indirect
processes. Because the UV absorption lines by which the
photodissociation is initiated can become optically thick, these
molecules can shield molecules lying deeper into the cloud through
`self-' or `mutual' shielding \citep{vanDishoeck88}.

Molecular ions are efficiently destroyed by the process of
dissociative recombination with electrons, which is very rapid at low
temperatures.  Rate coefficients in 10--30 K gas are typically
$10^{-7}$ to $10^{-6}$ cm$^3$ s$^{-1}$.  The dissociative
recombination of H$_3^+$, a key species in the chemistry, has been
subject to considerable discussion in the last three decades, but
experiments appear to converge on a rapid value of $\sim 10^{-6}$
cm$^3$ s$^{-1}$ at low temperatures \citep{McCall03}.

In contrast with the absolute rate coefficients, the branching ratios
to the various products are a major uncertainty. Experiments and
theory have produced different results. Experiments
indicate that three--body product channels (e.g., H$_3$O$^+$ + e $\to$
OH + H + H) have a much larger probability than thought previously.
The fraction of dissociations leading to the largest stable molecule
is often small. For example, the probability of forming H$_2$O in the
dissociative recombination of H$_3$O$^+$ is only 17\% \citep{Buhr10}
and that of CH$_3$OH in the recombination of CH$_3$OH$_2^+$ only
$\sim$6\% \citep{Geppert06}.

In dense clouds, destruction of neutral molecules can also occur
through chemical reactions (see below). The He$^+$ ion, formed by the
cosmic--ray ionization of He, is particularly effective in breaking
bonds since 24 eV of energy is liberated in its neutralization (e.g.,
He$^+$ + N$_2$ $\to$ He + N$^+$ + N or He$^+$ + CO $\to$ He + C$^+$ +
O).  Collisional dissociation of molecules is only important in
regions of very high temperature ($>$3000 K) and density such as
shocks in the vicinity of young stars.

\subsubsection{Rearrangement.}  Once molecular bonds have been formed, they can
be rearranged by chemical reactions leading to more complex
species. Initially mostly ion--molecule reactions were considered in
the models, primarily because the vast majority of them are very rapid
down to temperatures of 10 K \citep{Herbst73}. If the reaction is
exothermic, the simple Langevin theory states that the rate
coefficient is independent of temperature, and depends only on the
polarizability of the neutral molecule and the reduced mass of the
system, leading to typical values of $\sim 10^{-9}$ cm$^{-3}$
s$^{-1}$.  It was subsequently realized that reactions between ions
and molecules with a permanent dipole (e.g. C$^+$ + H$_2$O) may be
factors of 10--100 larger at low temperatures, because of the enhanced
long--range attraction.  The ions are produced either by
photoionization (C$^+$) or by cosmic rays producing H$^+$ or H$_2^+$.
H$_2^+$ reacts quickly with H$_2$ to form H$_3^+$, whereas H$^+$ can
transfer its charge to species like O.  These ions subsequently react
rapidly with neutral molecules down to very low temperatures, as long
as the reactions are exothermic and have no activation barrier.

Over the last two decades, experimental work has demonstrated that
radical--radical (e.g. CN + O$_2$) and radical--unsaturated molecule
(e.g. CN + C$_2$H$_2$) reactions have rate coefficients that are only
a factor of $\sim$5 lower than those of ion--molecule reactions at low
temperatures. Even some radical-saturated molecule reactions can occur
in cold clouds (e.g. CN + C$_2$H$_6$).  Also, reactions between
radicals and atoms with a non--zero angular momentum (e.g.
O($^3$P$_2$) + OH) are fast at low temperatures \citep{Atkinson04}.

In deciding which reactions are most important in the formation of a
certain species, a few simple facts should be kept in mind.  First,
the abundances of the elements play an important role (see
Table~\ref{tab:abundances}).
Because hydrogen is so much more abundant than any other element,
reactions with H and H$_2$ dominate the networks if they are
exothermic. This is only the case for small ions.  Most reactions of
neutrals and large ions with H or H$_2$ have substantial energy
barriers or are endothermic, and therefore do not proceed at low
temperatures. Reactions with the next most abundant species then
become important, especially with ions because of their large rate
coefficients.  The ions are produced either by photoionization by
the ISRF at the edges of clouds (e.g., C$^+$) or by cosmic rays deep
inside clouds (e.g., H$_3^+$).
Because the ionization potentials of O and N are larger than 13.6 eV,
these elements cannot be photoionized by the ISRF and are therefore
mostly neutral in interstellar clouds.

Compilations of reaction rate coefficients together with codes that
solve the coupled differential equations include the UMIST 2013
database \citep{McElroy13} and the KIDA database \citep{Wakelam12}
\footnote{{\tt www.adfa.net} and {\tt kida.obs.u-bordeaux1.fr}}. The
latter website includes the Nahoon (formally known as Ohio State)
gas-grain chemistry code by Herbst and co-workers.

\subsection{Gas-grain chemistry models}
\label{sect:gasgrain}

\begin{figure}[h]
\centering
  \includegraphics[width=8cm]{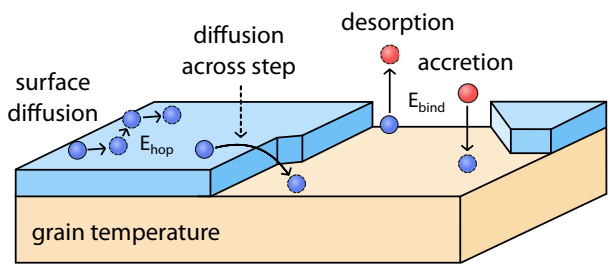}
  \caption{Illustration of different binding sites on a rough
    surface. Figure by M.\ Persson. }
  \label{fig:barriers}
\end{figure}

\subsubsection{Surface chemistry.}
\label{sect:surface}
The overall efficiency of surface reactions depends on
the probability that the atoms or molecules stick to the grains upon
collision, their mobility on the surface, the probability that
molecule formation occurs, and finally the probability that the
molecule is released back into the gas phase.  

The bulk of the surface reactions are assumed to proceed by diffusion
of at least one of the two reactants on the surface to find the other
partner. This requires a prescription of the diffusion of the reactant
to hop from one surface site to another. Usually, this is described by
a standard reaction rate coefficient $K_{\rm hop}=\nu \ {\exp(-E_{\rm
    hop}/kT_s})$ where $\nu$ is the vibrational frequency of the
reactant to the surface, $E_{\rm hop}$ is the energy barrier to hop
from one site to another and $T_s$ is the surface temperature. In most
models, this barrier is taken to be a constant fraction of the binding
energy, $E_{\rm hop}=c^{\rm st}\times E_{\rm bind}$ with $c^{\rm st}$
varying from 0.3 to 0.7 in different models \citep{Aikawa01}. However,
this prescription does not take into account that the surface is rough
and that the hopping barriers change from site to site
\citep{Cuppen07} (see Fig.~\ref{fig:barriers}). Also, the importance
of tunneling at the lowest temperatures is still debated and the
formulation for the competition between diffusion and reaction is not
clear. 

The competing Ely-Rideal mechanism, in which an atom or molecule from
the gas lands directly on top of the reactant on the surface can also
play a role under some conditions, as do `hot atom' reactions in which
the atom lands on the surface with excess energy that can be used to
overcome barriers. Regardless of the precise process, it is clear that
surface temperature plays a critical role in the ability of species to
react.  Light species such as H, H$_2$, C, N and O can likely hop over
the surface to find a reaction partner even at $T_D\approx 10$~K,
whereas heavier species are immobile at low dust temperatures.
Reactions of these atoms with CO can produce a number of `cold'
complex molecules (see \S~\ref{sect:hotcores}) \cite{Tielens97}.

Chemistry occurs not only on the surface of the grains but also deep
inside the ice. UV photons can penetrate at least 50 monolayers and
dissociate molecules both on top and deep inside the ice.  The
resulting atoms and radicals are initially highly mobile because of
the excess energy from the dissociation process but quickly loose this
energy on pico-second timescales and become
trapped\citep{Andersson06}. A second route to producing
complex organic molecules invokes that these radicals become mobile and
find each other once the ice temperature increases from $\sim$10 K
to 20--40 K \citep{Garrod06,Garrod08}.

Alternatively, cosmic rays can penetrate ices and create a wealth of
chemical complexity, and there is a rich literature on laboratory
experiments bombarding ices with high energy particles
\citep{Moore01,Islam14,Kaiser11}. The relative importance of UV vs
cosmic ray processing in creating complex organic molecules is still
under discussion. \citet{Shen04} have argued that UV radiation is more
important since it deposits 10$\times$ more energy per molecule in the
ice than cosmic rays for typical cosmic ray fluxes (see their Table
3). On the other hand McCoustra argues in this volume that each cosmic
ray triggers multiple events compensating for the lower energy
deposition rate. Overall, the chemical consequences of UV and
high-energy particle processing of ices may be rather similar
\citep{Munoz14}, although the very strong CO and N$_2$ bonds are
usually not broken in the UV irradiation case.  A new aspect discussed
at this conference in papers by Mason, Boamah, Siemer and Maity et
al.\ is to what extent electron bombardment and charged species in
ices can affect the production of complex molecules.

While there is a growing set of laboratory experiments studying
reactions in ices of the types described above, it is not yet clear
how to translate the laboratory data to astronomical model
parameters. Timescales are also an issue: individual processes occur
on picosecond timescales but lab experiments usually measure changes
of bulk ice as function of fluence over a period of
hours. Astronomical applications involve timescales of $> 10^5$ yr,
and further modeling is needed to translate laboratory results to
these different regimes \citep{Cuppen09,Cuppen13}.

\subsubsection{Gas-grain interactions.}
The standard treatment of the interplay between gas phase and grain
surface chemistry is through rate equations \citep{Hasegawa93}, the
basics of which are described in more detail elsewhere
\citep{Herbst13,vanDishoeck13}.  The standard rate equation approach
is known to be inadequate under some conditions, however, especially
for models with very small grains and only a few species per grain.
Many alternative approaches are being considered such as the modified rate
equations, Monte Carlo, Master equation, and hybrid methods, each with
their advantages and drawbacks.

The terms coupling the two regimes are accretion of gas-phase species
onto the grains and desorption of molecules from the grains back into
the gas. The accretion basically involves the rate of collisions of
the gas-phase species with the grain and the probability that the
species then sticks on the grain. This term is in principle
straight-forward to implement and depends on the geometrical surface
area of the grains as well as the sticking coefficient $S$ as function
of temperature. Usually, $S$ is taken to be 1 for heavy species at low
temperatures, as also suggested by laboratory experiments
\citep{Bisschop06}. For the important case of atomic H sticking to
ice, the temperature dependence of $S(T)$ has been computed
\citep{AlHalabi07} and is now also being measured
\citep{Matar10}. Older models have often adopted a value of $S$ less
than unity, say 0.8 or 0.9, as a `fudge factor' to implicitly prevent
all species from freezing out onto the grains within $10^5$ yr, rather
than treating the desorption processes explicitly.

In terms of desorption, there is a growing list of mechanisms that can
return molecules from the ice into the gas. These include (i) thermal
sublimation \citep{Collings04,Brown11}; (ii) UV photodesorption
\citep{Oberg09h2o}; (iii) cosmic-ray induced spot heating
\citep{Leger85}; (iv) cosmic-ray whole grain heating \citep{Leger85};
(v) exothermicity from chemical reactions \citep{Garrod07,Dulieu13};
and (vi) ice mantle explosions \citep{Hendecourt82}.  Process (i) is
well measured in the laboratory for many species but
is clearly not operative in cold dark clouds where dust temperatures
are only 10 K. Process (iii) may be effective for weakly bound
molecules like CO but not for strongly bound species like H$_2$O,
whereas process (iv) is generally negligible except for the smallest
grains. Processes (v) and (vi) may contribute but are still poorly
characterized experimentally.

This leaves process (ii), UV photodesorption, as a prime non-thermal
mechanism for getting molecules off the grains in cold dense
clouds. The UV radiation is provided not only by the external ISRF but
also has contributions from the internally produced UV photons by the
interaction of cosmic rays with H$_2$ (\S~1). Thanks to a series of
molecular dynamics simulations \citep{Andersson06,Arasa10} and
laboratory experiments \citep{Oberg09h2o,Fayolle11}, the
photodesorption yields are now being quantified for the main ice
species like CO and H$_2$O (see also Fillion et al., this volume). The
process is induced by absorption into excited electronic states
(either directly dissociative or indirectly through an exciton state),
so the photodissociation cross section or yield is wavelength
dependent just as for gas-phase molecules. It is also possible that UV
excitation of one molecule kicks out a neighboring molecule, as
demonstrated experimentally for the case of N$_2$ by CO
\citep{Oberg07,Bertin13}.

Basic input parameters for any gas-grain model are the overall
elemental abundances, the initial molecular abundances at $t=0$ in the
case of time-dependent models, the primary cosmic ray ionization rate
$\zeta_{\rm H}$ and the grain size distribution.  The models then
provide number densities or concentrations $n$ (in cm$^{-3}$) of a
certain molecule as function of time and/or position within a
cloud. The fractional abundance of the molecule AB with respect to
H$_2$ is given by $n$(AB)/$n$(H$_2$).  Observers measure column
densities in cm$^{-2}$, i.e., the number density $n$ in cm$^{-3}$ of a
species integrated along a path, $N=\int n \ {\rm d}z$. An empirical
relation between extinction and the column density of hydrogen nuclei
has been found \citep{Bohlin78,Rachford09} $N_{\rm
  H}=N$(H)+2$N$(H$_2$)=$1.8\times 10^{21}$ $A_V$ cm$^{-2}$, which is
often used to present model results as a function of $A_V$ rather than
pathlength $z$. For diffuse clouds or PDRs proper integration along
the line of sight is done, but for dark clouds local model
concentration ratios are usually compared with observed column density
ratios.

\begin{figure}[h]
\centering
  \includegraphics[width=10cm]{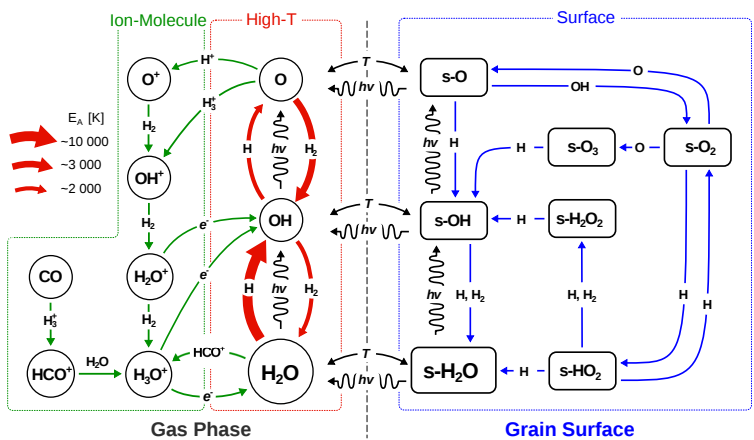}
  \caption{Simplified water chemistry illustrating the different routes to water through (left) low temperature gas-phase ion-molecule chemistry; (middle) high temperature gas-phase chemistry; and (right) surface chemistry. Figure by M. Persson, based on \citet{vanDishoeck11} and \citet{vanDishoeck13}.}
  \label{fig:waterchem}
\end{figure}

\subsubsection{Water as an example.}
An illustrative example of how different routes contribute to the
formation of a particular molecule under different conditions is
provided by the networks leading to interstellar water
\citep{vanDishoeck13} (Fig.~\ref{fig:waterchem}). At low temperatures
and densities, the ion-molecule reaction route dominates and produces
a low fractional abundance of water around $10^{-7}$. At high
temperatures such as encountered in shocks, reaction barriers of O and
OH with H$_2$ can be overcome and H$_2$O is rapidly formed by
neutral-neutral reactions. Finally, in cold dense clouds, formation of
water ice is very efficient and locks up the bulk of the oxygen not
contained in CO. Water can be brought from the ice into the gas phase
by photodesorption in cold clouds and by thermal desorption at high
temperatures.

\subsection{Model results: managing expectations and `back to basics'}

The reliability of the results of any chemical model depend on the
accuracy of the rate coefficients of the hundreds or thousands of
reactions contained in the databases. The rate coefficients, in turn,
are provided by chemical physics experts carrying out the relevant
experiments or calculations.  Ideally, each rate coefficient in a
database should be motivated by a critical evaluation of the chemistry
literature on that particular reaction by an independent expert: the
latest measurement is not necessarily the best or the most appropriate
value for astrochemical applications. Also, often extrapolations
beyond measured temperature regimes have to be made and
motivated. This is a time consuming process and one that is not highly
valued by funding agencies nor by universities in terms of career
paths.

It is therefore important to recognize that astrochemical databases
are put together and updated on a `best effort' basis. Following the
good example of atmospheric chemistry, where a two decade long effort
has resulted in a set of critically evaluated and motivated rate
coefficients \citep{Atkinson04}, the KIDA database has an (estimated)
uncertainty associated with each rate coefficient which can then be
propagated in the network. \citet{Wakelam10b} show examples of such
sensitivity analyses for a pure gas-phase chemistry network,
demonstrating that even the abundances of simple molecules like
H$_2$O, SO or CH, have an uncertainty of a factor of $\sim$3 just from
the uncertainties in individual rate coefficients. For larger
molecules, the cumulative effect of uncertainties in many more rates
easily results in an order of magnitude overall uncertainty. Such
differences are comparable to those found when two independent
networks are run for the same physical model.

So what constitutes good agreement between models and observations?
For diffuse clouds, the column densities of simple molecules like CH,
C$_2$, CN, OH, H$_2$O and HCl can be reproduced within a factor of 2
or better if the physical conditions of the cloud are independently
constrained \citep{vanDishoeck98ma2,Hollenbach12,Flagey13}. Even for
such relatively simple clouds with well constrained temperatures and
densities, there are well-known exceptions like CH$^+$ that require
different physical processes to be added to the model such as
turbulence (see \S~\ref{sect:diffuse}).  Nevertheless, diffuse clouds
and PDRs are still he best laboratories for `precision
astrochemistry'.

For dense clouds, the agreement between models and observations is
generally much worse. The chemically-rich dark cloud TMC-1 has served
as one of the main testbeds for decades and here agreement within an order of
magnitude for 80\% of the observed species is considered good
\citep{Terzieva98}. In spite of considerable laboratory effort, there
has not been much progress in this comparison over the past few
decades. Should we aim to do better as a community?  Or is this
uncertainty the inevitable consequence of the cumulative effect of
uncertainties in many individual rate coefficients with little hope
for improvement? Or do these dicrepancies point to other ingredients
that are missing in the models, such as the physical and dynamical
evolution of the source or small scale unresolved structure?

For this reason, several groups have started to go `back to basics'
and to ask more specialized questions that can be addressed with more
limited networks in order to isolate the critical chemical processes
at work.  For example, detailed models of well characterized dark
cores like B68 or L1544, for which the physical structure is well
determined independently, demonstrate that impressive agreement with
observations can be achieved for selected species, even on a linear
scale \citep{Tafalla02,Bergin02,Keto10,Padovani11}. Similarly, the
water abundance profile in low-mass pre- and protostellar cores is well
described by a simple network \citep{Caselli12,Mottram13,Keto14,Schmalzl14}.

This leads to two different approaches for modeling observational data
\citep{Doty04}. On the one side is the `forward model' in which a
full-blown chemical model is applied to a physical structure of the
source and the output is then compared with observations. Such a model
can turn many knobs to improve comparison with the data and the
question is what one has learned if good agreement is finally
obtained. The alternative approach is the `backward' or `retrieval'
model in which a trial abundance of a molecule is taken within a
physical model of the source, and this abundance is then varied to
obtain best agreement with observations. A more sophisticated case of
`retrieval' is where the trial abundance is inspired by the full
chemical models (e.g., by applying abundance jumps at certain
locations or adopting a functional form). The hybrid `back to basics'
method described above uses a simple or minimal chemical network of
the particular molecule under study to isolate the principal chemical
routes.  As deeper and higher angular resolution data are becoming
available with ALMA, this approach of addressing more focused
questions may be key to progress in the field.

In the following, major developments in observations and models of
specific regions in the lifecycle from clouds to planets are
discussed.

\section{Diffuse clouds and dense PDRs}

\subsection{Diffuse and translucent clouds}
\label{sect:diffuse}

The study of diffuse and translucent clouds with visual extinctions of
a few mag has obtained a large boost with the recent {\it
  Herschel}-HIFI data \citep{Gerin10}.  Rotational lines of molecules
are seen in absorption in diffuse clouds throughout the Galaxy along
the lines of sight toward distant far-infrared sources. Usually only
the lowest $J= 1-0 $ line is observed because higher levels do not
have sufficient population under these tenuous conditions. These
far-infrared data complement continued studies of other molecules like
H$_2$, C$_2$, C$_3$ which can only be observed by traditional optical
and UV absorption lines toward bright stars
\citep{Rachford09,Schmidt14}. Only a few molecules, most notably CH
and CH$^+$, can be observed by both techniques, although not for the
same lines of sight.

Because H$_2$ cannot be observed directly toward far-infrared sources,
the use of other molecules as tracers of H$_2$ have been
investigated. CH is a good candidate \citep{Gredel93}, but one of the
best options has proven to be the HF molecule
\citep{Neufeld10hf,Sonnentrucker10}. Even though the elemental
abundance of fluor is only $4\times 10^{-8}$, the fluorine chemistry
is particularly simple because the F + H$_2$ $\to$ HF + H reaction is
exothermic and has been very well studied by the chemical physics
community. The reaction of HF with C$^+$ produces CF$^+$ which has
also been detected, further confirming its chemistry.

A summary of the state of the observations and models prior to {\it
  Herschel} has been given in a number of reviews
\citep{vanDishoeck98ma2,Snow06}. The recent detections of OH$^+$ and
H$_2$O$^+$ point toward the existence of a new type of diffuse clouds
with only a small molecular fraction\citep{Gerin10}. The H$_2$/H ratio
must be low, of order 10\%, because both ions react rapidly with H$_2$
to form eventually H$_3$O$^+$. This mixed H/H$_2$ phase must be
ubiquitous in the interstellar medium since it is seen in absorption
along nearly every line of sight in the Galaxy, and even in emission
throughout entire galaxies in the local and distant Universe
\citep{vanderWerf10,Weiss13}.  For the same reason, ArH$^+$ probes
nearly pure atomic hydrogen gas, with an even lower H$_2$/H ratio of
$10^{-4}$.

The long-standing puzzle of the high abundance of CH$^+$ in diffuse
clouds, first detected nearly 80 years ago \citep{Douglas41}, is still
not fully solved, although models of the chemistry in turbulent
clouds, where the dissipation of turbulent energy provides the heat to
overcome the C$^+$ + H$_2$ endothermicity of 4640 K, is a plausible
explanation \citep{Godard09,Falgarone10}. The physics of interstellar
turbulence is not yet understood from first principles, however, and
non-Maxwellian motions between ions and neutrals need to be taken into
account in the presence of magnetic fields.  Similarly, the formation
of SH$^+$, detected both with ground-based telescopes and with {\it
  Herschel}-HIFI, requires extra energy for the S$^+$ + H$_2$ reaction
(+9860 K) to proceed \citep{Menten11,Godard12}. Both species have
somewhat larger line widths than those of CH and CN, $\sim$4 vs
$\sim$2 km s$^{-1}$, further justifying that additional physical
processes need to be invoked to drive the reactions.

\subsection{Photon-dominated regions}
\label{sect:pdr}
 
Photon-dominated or photodissociation regions (PDRs) are dense clouds
exposed to intense UV radiation, which controls both the chemistry and
heating of the gas. They are the high $I_{\rm UV}$ ($>10^3$), high
$n_{\rm H}$ ($> 10^4$ cm$^{-3}$) versions of the diffuse clouds
discussed above. At the edge of the cloud, the gas temperature becomes
so high ($\sim 1000$~K) that endothermic processes and reactions with
energy barriers can proceed, most notably the C$^+$ + H$_2$ $\to$
CH$^+$ + H and the O + H$_2$ $\to$ OH + H reactions
\citep{Sternberg95}. Subsequent reactions with C$^+$ lead to other
characteristic PDR tracers such as CO$^+$.

The Orion Bar PDR and the Horsehead nebula are prototype examples of
dense PDRs which have been studied in great detail observationally.
The Orion Bar is the clearest case exhibiting the layered chemical
structure: C$^+$ and PAH emission (pumped by UV) is seen close to the
star, followed by warm H$_2$ emission and then cool CO
\citep{Tielens93}. This structure is also reflected in the chemistry
of other species, with radicals such as C$_2$H, C$_4$H and C$_3$H$_2$,
whose formation benefits from the presence of free atomic carbon,
peaking ahead of molecules like C$^{18}$O
\citep{Hogerheijde95,vanderWiel09,Pety05,Gerin09}. Deep searches are
now starting to reveal more complex organic molecules in PDRs, mostly
those belonging to the `cold' complex type (see
\S~\ref{sect:surface} and \ref{sect:hotcores}) (Guzm\'an et
al, this volume).

Because of exposure to intense radiation and high temperatures,
excited ro-vibrational levels of molecules can be pumped. The
resulting populations are large enough that {\it state-to-state} processes
become significant in dense PDRs. The most prominent example is that
of CH$^+$ formation through reactions of C$^+$ with H$_2$(v,$J$), where
the H$_2$ ro-vibrational levels are pumped by UV radiation. These
state-specific reactions were included in PDR models developed in the
1980s and 1990s but have recently been revived in the context of dense
PDRs and the surface layers of protoplanetary disks
\citep{Agundez08,Thi11,Godard13}.

A second example is {\it formation pumping}, in which the reaction produces
a molecule in an excited ro-vibrational level which then radiates
before it collides with H$_2$. This process must dominate the
excitation of those species that react on every collision
\citep{Black98}. Examples are C$^+$ + H$_2$ $\to$ CH$^+$(v,$J$) + H
\citep{Godard13}, C$^+$ + OH $\to$ CO$^+$($J$) + H \citep{Stauber09}
and the processes leading to the formation of H$_3$O$^+$ ($J$) 
\citep{Lis14}.

Finally, excitation by electrons rather than H$_2$ or H can become
significant in PDRs. It explains why the HF line can be seen in
emission in the Orion Bar rather than absorption \citep{vanderTak12}.

\section{Cold molecular cores}
\label{sect:cold}

\subsection{H$_2$D$^+$ and extreme deuteration}
\label{sect:deuteration}

Cold molecular clouds have long been known to harbor high abundances
of deuterated molecules such as DCO$^+$ and DCN \citep{Wootten87},
with DCO$^+$/HCO$^+$ and DCN/HCN ratios at least three orders of
magnitude higher than the overall [D]/[H] ratio of $\sim 2\times
10^{-5}$ .  More recently, even doubly- and triply-deuterated
molecules such as D$_2$CO and ND$_3$ have been detected
\citep{Ceccarelli14}.  The observed ND$_3$/NH$_3$ ratio is about
$10^{-3}$, indicating an extreme deuterium enhancement of a factor of
$\sim 10^{12}$.

This huge fractionation has its origin in two factors. First, the
zero-point vibrational energy of deuterated molecules is lower than
that of their normal counterparts because of their higher reduced
mass. This makes their production reactions exothermic.  In cold
cores, most of the fractionation is initiated by the H$_3^+$ + HD
$\to$ H$_2$D$^+$ + H$_2$ reaction which is exothermic by about 230 K.
H$_2$D$^+$ then transfers a deuteron to CO or N$_2$ or another species.
Proof of this mechanism comes from direct observations of H$_2$D$^+$
in cold clouds \citep{Stark99,Caselli03}.  Even D$_2$H$^+$ has been
observed toward these sources \citep{Vastel04,Parise11}. Correct
calculation of their abundances requires explicit treatment of the
nuclear spin states (ortho and para) of all the species involved in
the reactions \citep{Pagani92,Pagani13,Walmsley04,Sipila10}.

The extreme fractionations observed in some clouds require an
additional explanation beyond just gas-phase reactions.  In the centers
of cold cores, there is now convincing observational evidence that
most of the heavy elements, including CO, are frozen out onto the
grains (see \S~\ref{sect:ices}).  Since CO is the main destroyer of
both H$_3^+$ and H$_2$D$^+$, their abundances are even further
enhanced when CO is removed from the gas. Indeed, HD then becomes the
main reaction partner of H$_3^+$ and the chemistry rapidly proceeds to
a highly deuterated state in which even D$_3^+$ can become comparable
to H$_3^+$ in abundance \citep{Roberts03}.  At slightly elevated
temperatures, the CH$_2$D$^+$ ion, formed by reaction of CH$_3^+$ +
HD, becomes more effective in controlling the DCN/HCN ratio,
consistent with observations of somewhat warmer PDRs and
protoplanetary disks \citep{Parise09,Oberg12}.

H$_3^+$ and HCO$^+$ are among the most abundant ions in dark clouds
and thus set the level of ionization in the cloud.  This, in turn,
controls the coupling of the gas to the magnetic field which slows
down the collapse of the cloud.  Since H$_3^+$ cannot be observed
directly in these clouds, the DCO$^+$/HCO$^+$ abundance ratio is
commonly used to infer the ionization fraction of cold cores. Typical
values are $ 10^{-9}$ -- $ 10^{-8}$ in the densest regions
\citet{Bergin07}.

\subsection{H$_2$O, O$_2$ and the importance of solid-state chemistry}
\label{sect:coldwater}

In the shielded regions of cold cores, the bulk of the heavy elements
are in ice mantles (see \S~\ref{sect:ices}). The laboratory
confirmation of the low temperature production of H$_2$O ice through
hydrogenation of O, O$_2$ and O$_3$ forms a highlight of solid-state
astrochemistry
\citep{Miyauchi08,Ioppolo08,Cuppen10,Dulieu10,Oba12,Chaabouni12}. In
low density regions the route through atomic O dominates, but at
higher densities the O$_2$ route becomes more important, with O$_2$
formed on the grains rather than frozen out from the gas.  This simple
chemistry reproduces well the {\it Herschel}-HIFI observations of
water in pre-stellar cores \citep{Caselli12} and protostellar
envelopes \citep{Mottram13}, provided that cosmic-ray induced UV
photodesorption is included in the models to get water off the grains
in the central part pf the core.

The O$_2$ route forms solid HO$_2$ and H$_2$O$_2$ as intermediates.
The success story of combined laboratory and observational studies
became complete with the detection of gas-phase HO$_2$
\citep{Bergman11} and H$_2$O$_2$ \citep{Parise12} with abundances
consistent with their grain surface formation \citep{Du11}. However,
both species have so far been detected in only one cloud, $\rho$ Oph
A, with searches in other clouds unsuccessful (Parise et al., this
volume).

Interestingly, $\rho$ Oph A is also one of only two positions where
interstellar O$_2$ gas has been firmly detected through multi-line
observations with the Odin and {\it Herschel} satellites
\citep{Larsson07,Liseau12}. Deep limits in other cold cores confirm
the scenario in which most of the O and O$_2$ must be converted into
H$_2$O ice on the grains before the protostar starts to heat its
surroundings \citep{Bergin10,Yildiz13o2}.  Perhaps the grain
temperature in $\rho$ Oph A, around 30 K, is just optimal to prevent
atomic oxygen from freezing out and being converted into water ice.  A
critical parameter in these models is the binding and diffusion energy
of atomic O to silicates and ices. The papers by Congiu, Lee and He
et al.\ discuss recent laboratory experiments which may suggest somewhat
higher values than used before\citep{Jing12}. They also highlight the
difficulties in providing a proper formulation for diffusion and
desorption of atoms at very low temperatures.

\begin{figure}[h]
\centering
  \includegraphics[width=9cm]{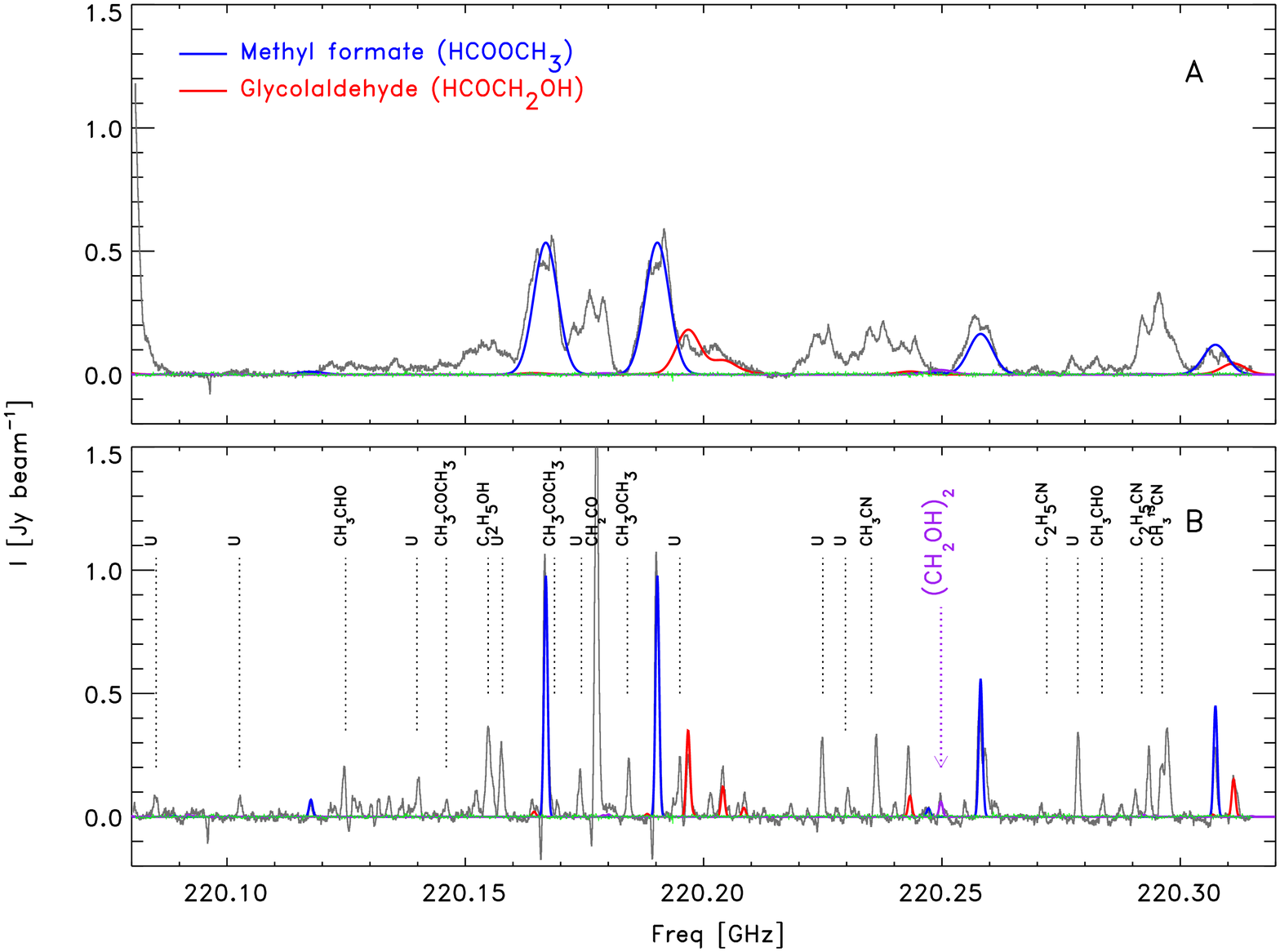}
  \caption{ALMA data toward the low-mass binary protostar
    IRAS16293A (upper) and IRAS16293B (lower) obtained with 16 antennas.  
    Fits from LTE models
    of the methyl formate (blue) and glycolaldehyde (red) emission are
    overplotted. The purple line indicates the model fit to the
    possible ethylene glycol transition. 
    Note the high density of lines, including several unidentified features,
   even in these early ALMA science data.  Reproduced with permission from
   \citet{Jorgensen12}.}
  \label{fig:i16293}
\end{figure}

\section{Protostars and hot cores: complex molecules}
\label{sect:hotcores}

Saturated complex organic molecules such as CH$_3$OCH$_3$, HCOOCH$_3$
and CH$_3$CN are prominently observed toward many (but not all)
high-mass protostars as well as toward some low-mass sources
\citep{Herbst09,Caselli12rev} (Fig.~\ref{fig:i16293}).  Three
generations of complex molecules are distinguished \citep{Herbst09}:
zeroth generation ice formation in cold dense cores prior to star
formation, which leads primarily to CH$_3$OH through hydrogenation of
CO. First generation organics are then formed during the subsequent
protostellar warm-up phase when radicals produced by photodissociation
of species like CH$_3$OH become mobile in and on the ices (first
generation, e.g., HCOOCH$_3$).  Second generation organics result from
high temperature gas-phase reactions involving evaporated molecules
(Fig.~\ref{fig:iceevolution}).

As the protostar heats the envelope, molecules sublimate from the ice
mantles, starting with the most volatile molecules like CO and N$_2$
in the outer envelope, and followed by species with larger binding
energies like CO$_2$. Once the grain temperature reaches 100 K close
to the protostar, even the most strongly bound molecules like H$_2$O
and CH$_3$OH sublimate from the grains, including any minor species
that are trapped in their matrices. These evaporated species then
drive a high temperature gas-phase chemistry producing second
generation complex molecules for a period of $\sim 10^5$ yr after
sublimation \citep{Charnley92}. Recent determinations of reaction rate
coefficients and branching ratios for dissociative recombination show
that this gas-phase route is less important than thought before,
however (see \S~\ref{sect:processes}).

This scenario can soon be tested in great detail with high angular
ALMA observations which can spatially resolve the hot cores and map
the different molecules. A `sweet taste' of what to expect is the
recent detection of the simplest sugar, glycolaldehyde (HCOCH$_2$OH),
in the low-mass protostar IRAS 16293-2422B (Fig.~\ref{fig:i16293}).
The emission comes from a region of only 25 AU in radius,
i.e., comparable to the orbit of Uranus in our own solar system
\citep{Jorgensen12}. Its abundance with respect to related complex
molecules like methyl formate and ethylene glycol is consistent with
laboratory experiments of mild photoprocessing of methanol-rich ice
mantles \citep{Oberg09meth}, although other reaction routes such as
discussed by Boamah et al, this volume, are not excluded.

The above scenario requires that the grains spend some time at
elevated temperatures so that the radicals become mobile (see
\S~\ref{sect:gasgrain}).  The recent detection of some complex organic
molecules in very cold sources that have never been heated much above
10~K therefore came as a surprise
\citep{Arce08,Oberg10b1,Oberg11ser,Bacmann12}. Most of these species
are the same as those identified as `cold' complex molecules in the
survey of hot cores by \citet{Bisschop07}, i.e., molecules with
excitation temperatures below 100 K originating in the colder outer
envelope rather than the hot core. The papers by \"Oberg et al.\ and
Guzm\'an et al.\ image these molecules in protostellar
envelopes\citep{Oberg13irs9} and PDRs, respectively.  Possible
explanations that have been put forward include enhanced radiative
association rate coefficients for these species or using the
exothermicity of the reactions\citep{Vasyunin13}. Alternatively, low
temperature surface reactions of atomic C and H with CO can naturally
lead to these species, if the binding energy of C to ice is low enough
(see Guzm{\'a}n et al.). Similarly, reactions with N and H with CO
could lead to nitrogen-bearing species like HNCO and even NH$_2$CHO.

\begin{figure}[h]
\centering
  \includegraphics[width=11.5cm]{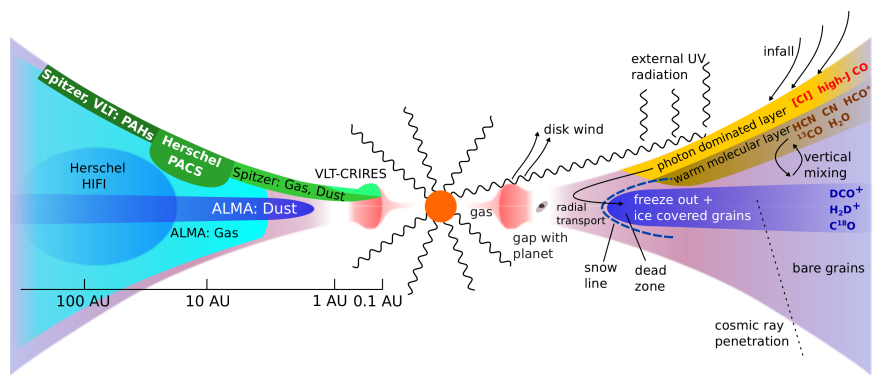}
  \caption{Cartoon illustration of the various physical and chemical processes taking place in protoplanetary disks (right) that are probed by different observational facilities (left). Figure by S. Bruderer.}
  \label{fig:disksketch}
\end{figure}

\subsection{Protoplanetary disks} 

Disks around young stars are the birthplaces of planets and are
therefore particularly important targets for astrochemistry. However,
disks are at least a factor of 1000 smaller than the clouds in which
they are formed and they contain only $\sim$1\% of the mass. Thus,
their emission is readily overwhelmed by that of any cloud and they
can only now be properly studied with the new generation of high
angular resolution and high sensitivity instruments. Disks are heated
by the radiation of their parent star so they have a radial and
vertical temperature gradient in both the gas and dust.  No single
instrument or wavelength probes the entire disk reservoir: a
combination of near-, mid-, far-infrared spectroscopy combined with
spatially resolved ALMA data is needed (Fig.~\ref{fig:disksketch}).
As a result of this physical structure, disks consist of different
chemical layers \citep{Aikawa02}: at the surface, molecules are
dissociated into atoms by the strong UV radiation. Deeper in the disk,
the grains are still warm enough to prevent freeze-out and molecules
are shielded enough from the UV to survive. Deep in the cold miplane
most molecules except H$_2$, H$_3^+$ and their isotopologs are frozen
out onto the grains.

There has been enormous observational and model activity in this field
in recent years, see recent reviews
\citep{Bergin07,Bergin13bio,Henning13,Pontoppidan14,Dutrey14}.
Highlights include the {\it Spitzer} and ground-based detections of
hot ($\sim 300-800$ K) C$_2$H$_2$, HCN, H$_2$O and CO$_2$ originating
in the inner $\sim$1 AU of disks
\citep{Lahuis06,Carr08,Salyk08,Pascucci09,Salyk11,Pontoppidan10,Najita13}.
CH$_4$ has been found in one disk \citep{Gibb13}, but no detection
of NH$_3$ has yet been reported \citep{Mandell12}. This rich array of
lines of simple molecules is primarily seen in disks around low-mass
stars that are cooler than the Sun; they are absent toward higher mass
stars with $T_{\rm eff}\approx 10,000$~K. This points again to the
importance of the wavelength dependence of the radiation field and
associated photodissociation \citep{vanDishoeck06}. The implications
for the elemental C and N budget in disks is discussed in the papers
by Bergin and Pontoppidan et al.\ in this volume.

Cooler H$_2$O and OH have been detected somewhat further out and
deeper into the disk with {\it Herschel}-PACS in a few sources
\citep{Fedele13,vanDishoeck14}. The bulk of the cold water reservoir
has been revealed by {\it Herschel}-HIFI in two disks. Because the
observed water gas is produced primarily by UV photodesorption of ice,
it points to the presence of a large reservoir of underlying water ice
\citep{Hogerheijde11}.

Another related highlight is the first imaging of `snow lines' in
disks, i.e., the radius where a molecule changes from being primarily
in the gas phase to being frozen out as ice. Because of the vertical
structure of disks, a `snow surface' is a more appropriate
description. Snow lines are important because they enhance the mass of
solids by a factor of a few and thus facilitate planet
formation. Also, the coating of grains with water ice enhances the
coagulation of grains, the first step in planet formation.  Because of
their lower binding energies, the snow lines of CO and CO$_2$ are
outside that of H$_2$O.  This selective freeze-out of major ice
reservoirs can change the overall elemental [C]/[O] abundance ratio in
the gas and thus the composition of the atmospheres of giant planets
that are formed there \citep{Oberg11}. The CO snowline has been imaged
with ALMA through N$_2$H$^+$ observations in the nearby TW Hya disk
\citep{Qi13}. N$_2$H$^+$ is enhanced when its main destroyer, CO,
freezes out. Another example is the CO snow line in the HD 163296 disk
imaged in CO isotopologs and in the DCO$^+$ ion, whose abundance peaks
when CO freezes out (see \S~\ref{sect:deuteration})
\citep{Mathews13,Qi11}.

In contrast with these simple species, complex molecules have not yet
been detected in disks, not even methanol. The most complex molecules
found so far are H$_2$CO, HC$_3$N and C$_3$H$_2$. One option is that
the strong UV radiation in disks prevents the build up of more complex
species, although methanol should be detectable \citep{Walsh14}, see
Walsh et al. this volume.

Finally, transitional disks which have a hole or gap in their dust
distribution are a hot topic since these gaps are likely caused by
planets currently forming in the disk
(Fig.~\ref{fig:disksketch}). ALMA images of several transitional disks
show remarkably asymmetric dust structures, pointing to traps induced
by the young planets in which the mm-sized grains are collected
\citep{vanderMarel13,Casassus13,Perez14}. The survival of molecules
like CO and the chemistry of other species in dust-free cavities in
transitional disks has been modeled by \citet{Bruderer13}
\citep{Bruderer14} and is being tested against ALMA observations for
the case of H$_2$CO \citep{vanderMarel14}.  Molecules produced in ices
may be observable as well when the disk midplane is exposed to the
stellar photons at the outer edge of the cavity and molecules can be
desorbed \citep{Cleeves11}. Altogether, these data form a vivid
illustration and warning that gas and dust grains do not necessarily
follow each other but that they need to be treated separately in the
models.

\subsection{Exo-planetary atmospheres}

The detection and characterization of exo-planets is one of the
fastest growing fields in astronomy, see recent reviews
\citep{Seager10,Madhusudhan14}. A large number of transiting
exo-planets has now been detected, most recently with the {\it Kepler}
and {\it Corot} satellites, and their atmospheres are starting to be
studied through high precision measurements when the planet is in
front (primary transit) and behind (secondary eclipse) the
star. Detection of molecules like CO and H$_2$O have been claimed and
disputed, but both are now firmly identified in a few targets
\citep{Birkby13,Deming13}.  One of the main science goals of future
facilities is to study the composition of exo-planetary atmospheres,
not only of Jupiter-like planets but down to the (super-)Earth regime.
Chemical models of such atmospheres are actively being pursued using
techniques from astrochemistry \citep{Helling14,Agundez14}, coupled with
expertise from other fields such as aeronomy.

Liquid water is likely a prerequisite for the emergence of life and
much of the water in our oceans on Earth plausibly comes from the
impact of asteroids and comets containing ice. The water molecules
themselves are mostly formed on the surfaces of grains in the cold
interstellar clouds prior to collapse as in \S~\ref{sect:coldwater}.
Following the water trail from dense cloud cores through collapsing
envelopes to planet-forming disks and exoplanets is a major goal of
modern astrochemistry and has recently been summarized
\citep{vanDishoeck14}.

\section{Concluding remarks}

``Is astrochemistry useful''? The answer to this question, asked first
by Dalgarno in 1986, is a convincing ``yes'', with astrochemistry now
firmly integrated into astronomy and at the same time stimulating
chemical physics.  Indeed, the interaction between dust, ice and gas
has stimulated the development of solid-state astrochemistry and
provided new insights in basic chemistry. Continued close interaction
on well-defined questions that can be addressed by the combination of
laboratory-observations-models is needed.

There has been a clear shift from pure gas-phase chemistry to a
gas-dust-ice chemistry over the past decades, making the topic of this
Faraday Discussion very timely. The solid-state routes to water and
methanol, two key components of ices, have been firmly established
through laboratory experiments and observations, including detection
of the intermediates in the network. Gas-grain chemistry now has
predictive power and is likely at the basis of many of the complex
organic molecules seen in star-forming regions. However, the precise
routes are not yet understood, and the presence of the `cold' complex
molecules challenges the convential theory that the grain temperature
needs to be elevated above 10 K to drive the reactions. Other puzzles
to be solved include why some sources are more line-rich than others,
why oxygen- and nitrogen-containing complex molecules are sometimes
spatially separated, and why protoplanetary disks are poor in complex
molecules.

Because interstellar clouds and protostars are dynamic entities, there
is increased activity to couple the full gas-grain chemical models
with hydrodynamical models of cloud formation and collapse. Computing
power is now available to do so and impressive simulations are being
performed. An alternative approach is to couple chemistry with
semi-analytical models of protostar and disk
formation\citep{Visser11}. Other factors such as the detailed 3D
geometry of the source, dust evolution and gas-dust separation now
also need to be taken into account. While these avenues clearly need
to be pursued, this paper has also advocated the `back to basics'
approach to elucidate the main chemical and physical processes at
play.

This review has focused primarily on small molecules and moderate size
complex organic molecules built through reactions in and on ices,
highlighting the gas-grain interactions. A different route to
molecular complexity starts with the much bigger PAH molecules and
carbonaceous material, which are known to exist throughout the
Universe. Through a series of UV and X-ray photolysis, radical
reactions and combustion-type chemistry, these big molecules can also
be transformed into smaller species and organic material that can be
incorporated in future solar systems \citep{Tielens13}.  Also, the
carriers of the omni-present Diffuse Interstellar Bands, discovered
nearly a century ago in diffuse clouds, are still a mystery, see
recent reviews \citet{Cami14}.  This alternative chemistry cycle could
be an appropriate topic for a future Faraday Discussion.

On the observational side, ALMA is {\it the} astrochemistry machine of
the future and is expected to bring advances in several areas
\citep{Herbst08}. It can search for new molecules at least a factor of
100 deeper than before, including prebiotic molecules such as
glycine. Equally important, it will survey many more sources than just
the `classical' Orion and SgrB2 regions used so far to hunt for new
molecules. ALMA has the spatial resolution to image the relevant
physical-chemical scales for the first time, such as thin PDR and
shock layers, and it can resolve the hot cores and outflow cavities
close to protostars. With its high sensitivity and broad frequency
range, it can also probe hotter gas and reveal the importance of
state-selective processes. Finally, astrochemical studies of
high-redshift galaxies are enabled by ALMA, at the level that was used
to study galactic clouds 30 years ago.

Besides ALMA, astrochemistry has other exciting new missions on the
horizon. The {\it Rosetta} spacecraft will rendez-vous with the comet
67P/Churyumov-Gerasimenko in summer 2014 and deploy a lander to
investigate its surface composition in late 2014. This mission will
give a big boost to studies of the connection between solar system and
interstellar material. The 6-m {\it James Webb Space Telescope} (JWST)
will be launched in late 2018 and will have near- and mid-infrared
spectroscopic instruments to survey ices and gases with unprecedented
sensitivity. Finally, the next generation of Extremely Large
Telescopes (ELTs) is starting to be built and will provide the
ultimate spatial and spectral resolution at optical- and near-infrared
wavelengths. Overall, there will be plenty of material for future
exciting new `plots' and `plays' in astrochemistry.

\section*{Acknowledgements}

The author is grateful to Magnus Persson (Fig. 2,5,6), Sandrine
Bottinelli (Fig.3) and Simon Bruderer (Fig. 8) for providing figures.
Astrochemistry in Leiden is supported by the Netherlands Organization
for Scientific Research (NWO), The Netherlands Research School for
Astronomy (NOVA), by a Royal Netherlands Academy of Arts and Sciences
(KNAW) professor prize, and by European Union A-ERC grant 291141
CHEMPLAN.

\footnotesize{
\bibliography{biblio_evd} 
\bibliographystyle{rsc}
}

\end{document}